\documentclass{article}

\PassOptionsToPackage{numbers, sort&compress}{natbib}
\usepackage[final,main]{neurips_2026}      
\usepackage{amsmath}
\usepackage{booktabs}
\usepackage{hyperref}

\usepackage[table]{xcolor}
\usepackage{wrapfig}
\usepackage{graphicx}
\usepackage{amssymb}
\usepackage{todonotes}
\usepackage{wrapfig}
\newcommand{\x}{\mathbf{x}}

\newcommand{\K}{\mathbf{k}}

\newcommand{\ift}{\mathcal{F}^{-1}}

\usepackage{graphicx}
\usepackage{subcaption}
\DeclareMathOperator{\argmin}{argmin}

\usepackage{hyperref}
\title{Coarse-to-Fine 3D MRI Reconstruction via Resolution-Agnostic  Neural Operators}

\DeclareRobustCommand{\authorblock}{%
\setlength{\tabcolsep}{5pt}%
\begin{tabular}{@{}cccc@{}}
Jiayun Wang$^{1}$\thanks{Equal contribution.} &
Ram Daftari$^{2,3*}$ &
Weijie Gan$^{4}$ &
Valentin Duruisseaux$^{2}$
\\[3pt]
Ruibo Wang$^{5}$ &
Armeet Singh Jatyani$^{2}$ &
Boris Bonev$^{6}$ &
Anima Anandkumar$^{2}$
\\[6pt]
\multicolumn{4}{c}{\footnotesize $^1$Georgia Tech \quad $^2$Caltech \quad $^3$IIT Kanpur \quad $^4$Washington University in St.\ Louis \quad $^5$TU Delft \quad $^6$NVIDIA}
\\[3pt]
\multicolumn{4}{c}{\tt\footnotesize pjwang@gatech.edu,\ ramdd24@iitk.ac.in,\ anima@caltech.edu}
\\[3pt]
\end{tabular}%
}
\author{\authorblock}

\begin{document}
    \maketitle

\begin{abstract}
Deep-learning-based 3D MRI reconstruction is limited in practice by the GPU memory and compute cost of full-volume training; 3D acquisitions are thus most often reconstructed by 2D models applied slice-by-slice and then stacked, which ignores through-plane (stacking-axis) structure. We reconstruct in full 3D and ask whether a model can be trained at low resolution (fast and memory-feasible) yet deployed at higher resolution. This coarse-to-fine strategy needs reliable cross-resolution performance, but standard networks are resolution-specific and degrade substantially when evaluated outside the training setting. We therefore use resolution-agnostic neural operators (NOs) and propose \emph{KIR-NO} (\emph{k}-space-to-Image Reconstruction Neural Operator), a discretization-agnostic framework for accelerated 3D MRI reconstruction. KIR-NO is trained with full backpropagation on low-resolution volumes and applied zero-shot to higher-resolution volumes at inference. 
Its core building block is a 3D neural operator based on discrete-continuous convolutions: convolutional filters are parameterized in a continuous basis  and sampled at arbitrary voxel resolutions while preserving the local inductive bias of convolution. KIR-NO integrates these learned operators with physics-based consistency: a $k$-space neural-operator head first refines the undersampled Fourier measurements, and image-space neural-operator cascades then refine the reconstruction through interleaved data-consistency updates. Across 3D knee and brain MRI benchmarks, KIR-NO improves over classical and CNN baselines: it improves over a state-of-the-art 3D CNN by {$2.2$}\,dB PSNR at half resolution under $4\times$ acceleration on SKM-TEA, with gains persisting at $8\times$ and $16\times$. Together, these results show that resolution-agnostic operators make coarse-to-fine 3D MRI reconstruction -- cheap low-resolution training with high-resolution deployment -- practical across anatomies. Our code is available at \url{https://github.com/neuraloperator/KIR-NO}.
\end{abstract}

    \begin{figure*}[!t]
    \centering
    \includegraphics[width=\linewidth]{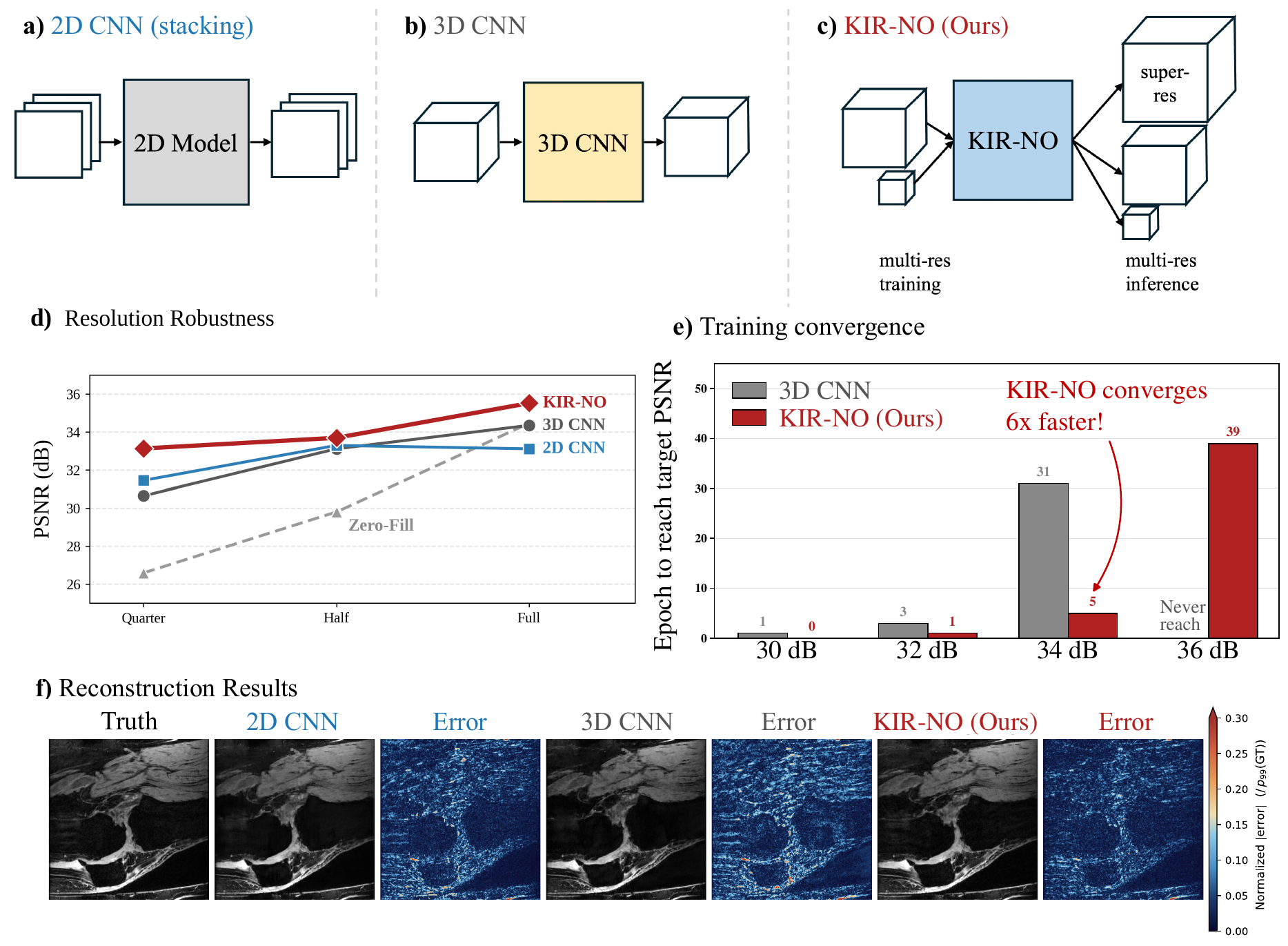}
    \caption{(a--c) Three deep learning architectures for 3D MRI reconstruction. \textbf{(a) 2D CNN:} GPU-memory limits force slice-wise 2D models that stack into a volume, losing inter-slice consistency. \textbf{(b) 3D CNN:} fixed discrete kernels are tied to the training grid, so cross-resolution transfer is unreliable. \textbf{(c) 3D Neural Operator (KIR-NO, ours):} discretization-agnostic --- trained on memory-friendly coarse volumes and deployed at higher resolutions without retraining. \textbf{(d) Resolution robustness:} PSNR at quarter/half/full resolution on SKM-TEA (quarter-trained); KIR-NO leads throughout. \textbf{(e) Convergence} (SKM-TEA, half-res, $4\times$): first epoch to reach each PSNR threshold; KIR-NO reaches $34$\,dB $\sim$$6\times$ faster than the 3D CNN. \textbf{(f) Qualitative:} a knee slice with error maps for 2D CNN, 3D CNN, and KIR-NO; KIR-NO has the lowest residual.}
        
    \label{fig:intro}
\end{figure*}

\section{Introduction}\label{sec:intro} 
Deep learning has emerged as a powerful tool for medical imaging, e.g. MRI reconstruction from highly undersampled measurements \citep{griswold2002generalized,lustig2008compressed,hammernik2018learning,e2evarnet,jatyani2025nomri,chung2022score}, where it can shorten scans and improve patient comfort. Although most existing work focuses on 2D reconstruction due to GPU memory and compute constraints~\citep{wang2021memory,wong2023fnoseg3d}, there is growing interest in 3D MRI reconstruction. Compared with slice-wise 2D approaches, 3D models leverage volumetric context to improve through-plane consistency and reconstruction fidelity~\citep{wang2021memory}, which is especially important when lesions span multiple slices and 3D morphology aids diagnosis~\citep{husband2001mri}. It is enabled by public volumetric datasets (e.g., SKM-TEA~\citep{desai2022skm}) and by hardware that increasingly makes whole-volume processing feasible.

The computational gap between 2D and 3D is significant at training time. While high-resolution 3D inference only holds forward activations, training additionally requires storing activations across all layers for backpropagation and computing gradients across an additional spatial dimension, making naive full-resolution training prohibitively expensive. This motivates a \textit{coarse-to-fine} paradigm: \emph{train at low resolution, where backpropagation remains feasible, and deploy zero-shot at full resolution.} The memory saving here is a property of the training \emph{recipe}, thus the decisive question is not model size but whether the learned model transfers reliably across resolutions. 
Standard CNNs, however, overfit to the discretizations seen during training~\citep{wong2023fnoseg3d}: resizing or interpolating discrete kernels alters the effective receptive field and provides no principled mechanism for resolution transfer. In MRI, such unprincipled transfer can introduce artifacts or hallucinated structures in critical anatomy and may mislead clinical interpretation.

Neural operators provide a principled alternative by learning function-to-function maps rather than mappings in fixed dimensions~\citep{azizzadenesheli2024neural,berner2025principled}, naturally supporting resolution-agnostic reconstruction. In particular, local neural operators built on discrete-continuous convolutions (DISCO)~\citep{ocampo2022,local_no} serve as a resolution-agnostic convolution layer, with strong empirical results on 2D MRI~\citep{jatyani2025nomri}, CT \cite{datta2025resolution} and 3D photoacoustic tomography~\citep{wang2025accelerating3dphotoacousticcomputed}, suggesting that they  adapt more naturally than standard convolutions for varying imaging resolution and undersampling.


\begin{figure}[t!] 
    \centering

    \includegraphics[width=\linewidth]{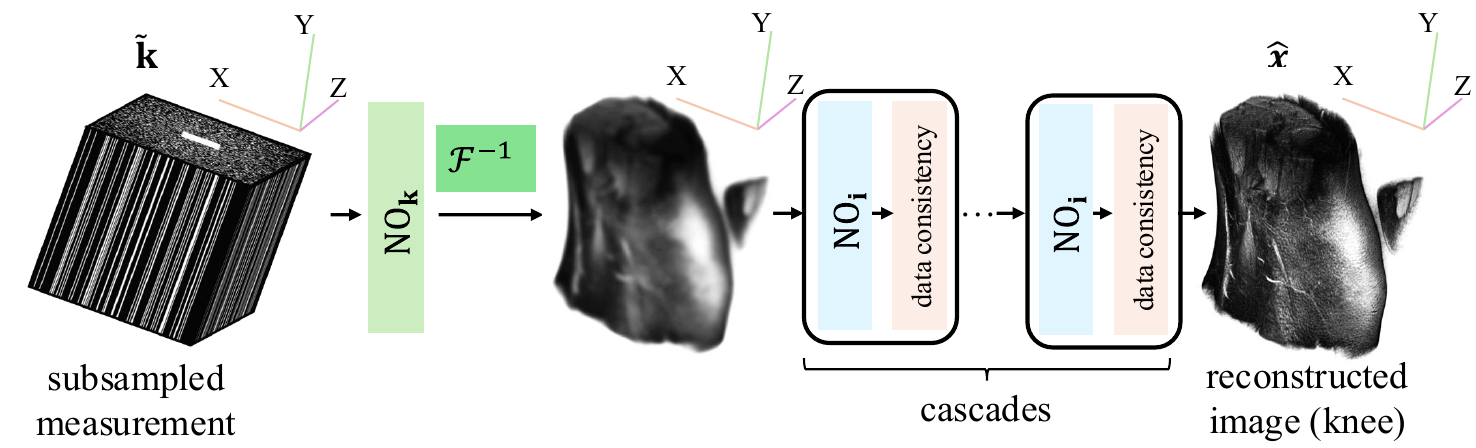} \\
    \caption{Our \textbf{KIR-NO} architecture. Undersampled multi-coil $k$-space measurements $\tilde{\K}$ are first refined by a $k$-space neural-operator head $\mathrm{NO}_k$ (Eq.~\ref{eq:kno_head}), then mapped to image space (inverse Fourier transform $\ift$ and coil combination) and refined by an unrolled cascade of image-space neural operators $\mathrm{NO}_i$ (Eq.~\ref{eq:unrolled_update}, where $\ell$ indexes the cascade) interleaved with data-consistency updates, producing the final 3D reconstruction $\hat{\x}$. Operator layers learn continuous kernels built on discrete-continuous convolutions, in contrast to fixed-resolution conventional CNN kernels.}
    
    \label{fig:arch}
\end{figure}

\hfill

\textbf{Our approach.}
Our central novelty is making resolution-agnostic reconstruction work \emph{natively in 3D}: while the prior 2D neural-operator/DISCO line targets robustness to undersampling patterns and field-of-view~\citep{jatyani2025nomri}, we target robustness to \emph{resolution}, a coarse-to-fine capability specific to 3D. Extending 2D DISCO to 3D is non-trivial: the 2D kernel bases~\citep{ocampo2022,local_no} are defined on planar supports and must be redesigned for volumetric kernels that stay compactly supported, so that each layer remains local and memory-feasible on dense 3D grids. We therefore design two new 3D continuous kernel bases --- piecewise-linear hat functions in spherical coordinates and 3D Morlet wavelets --- that are strictly local, orientation-selective, and GPU-friendly. Using these bases, we build \emph{KIR-NO}, a $k$-space 3D DISCO head refines the undersampled measurements, then (after an inverse FFT) an unrolled cascade of image-space 3D DISCO operators alternates with $k$-space data-consistency updates. Sampling each kernel at the target voxel resolution makes KIR-NO a resolution-agnostic alternative to \texttt{Conv3D} that preserves convolutional locality and weight sharing, leading to more reliable high-resolution transfer than 3D CNNs. 

Empirically across knee and brain MRI, KIR-NO outperforms classical and CNN baselines, and diffusion baselines given a single sampling step. On SKM-TEA it improves over a 3D E2E-VarNet by $+2.2$\,dB at the in-domain half resolution ($4\times$) and, most notably, by $+3$\,dB ($0.940$ vs $0.870$ SSIM) under zero-shot super-resolution to full resolution, with gains persisting at $8\times$ and $16\times$. 

\textbf{Contributions.} We (i) introduce a coarse-to-fine paradigm for accelerated 3D MRI enabled by resolution-agnostic operators: train at low resolution, deploy zero-shot at high resolution; (ii) design two new 3D DISCO kernel bases (piecewise-linear-spherical and 3D Morlet); and (iii) build \emph{KIR-NO}, a dual-domain $k$-space/image neural-operator reconstructor. The approach could extend to other volumetric inverse problems (e.g., seismic imaging, 3D microscopy).

    \section{Related Work} \label{sec: Related Work}

\textbf{Accelerated MRI.}
Medical imaging tasks are often inverse problems: recovering images from indirect, noisy measurements under a known forward model~\citep{lustig2007sparse}. In MRI, this model samples the image Fourier transform at discrete $k$-space frequencies; accelerated MRI undersamples these frequencies~\citep{lustig2008compressed}, making reconstruction ill-posed. This is especially challenging in 3D, where $O(n^3)$ voxels sharply increase memory and compute requirements~\citep{wang2021memory,desai2022skm}. MRI reconstruction is often cast as regularized optimization alternating between \emph{data consistency} and prior-driven regularization, using classical sparsity priors~\citep{chan2006total} or learned plug-and-play denoisers~\citep{gan2023block,kamilov2023plug}.

\textbf{Deep Learning for MRI Reconstruction.}
Early deep methods learn a feed-forward CNN that maps the zero-filled reconstruction to a clean image~\citep{sun2016deep,johnson2019conditional}. \emph{Unrolled} networks instead preserve the optimization structure by replacing the hand-crafted prior with a trainable module and unrolling iterations into \emph{cascades}, trained end-to-end~\citep{hammernik2018learning,e2evarnet}. Our main CNN baseline, E2E-VarNet~\citep{e2evarnet}, alternates multi-coil $k$-space data fidelity with U-Net image refinement and provided sensitivity maps. In 3D, storing per-cascade activations scales as $O(n^3)$, making full-volume training memory-prohibitive at clinical resolutions; many models therefore train on 2D slices and apply them slice-by-slice, sacrificing through-plane consistency. \citet{wang2021memory} address this with memory-efficient unrolled 3D training, but remain tied to a fixed training discretization, motivating our resolution-agnostic 3D framework. Score-based diffusion priors~\citep{chung2022score} decouple the prior from the forward operator but require many sampling steps and are substantially slower at inference.

\textbf{Neural Operators.} Unlike standard neural networks, neural operators learn function-to-function maps~\citep{azizzadenesheli2024neural,berner2025principled}, are universal approximators for nonlinear operators~\citep{Kovachki2021_2}, accept inputs and can be queried at arbitrary discretizations, enabling training across mixed resolutions. Physics-informed variants~\citep{li2024pino,lin2025mGNO,ganeshram2025fcpinohighprecisionphysicsinformed} add residual losses from governing equations; in contrast, KIR-NO incorporates MRI acquisition physics through data-consistency and $k$-space processing rather than an additional loss. The Fourier neural operator (FNO)~\citep{fnop,duruisseaux2025FNOGuide} truncates high-frequency modes and is equivalent to a global convolution; FNOSeg3D~\citep{wong2023fnoseg3d} adapts it for 3D MRI segmentation. Reconstruction, however, requires capturing intricate anatomical structure with local inductive biases, where FNO's high-frequency truncation is detrimental. DISCO~\citep{ocampo2022,local_no} retains all frequencies and learns local biases, and \citet{jatyani2025nomri} show DISCO-based operators are robust for 2D MRI across undersampling patterns and extended field-of-view. That line, however, is confined to 2D and targets robustness to the \emph{sampling} operator; we instead target cross-\emph{resolution} transfer in \emph{3D} (coarse-to-fine, train-low/deploy-high), which raises new continuous-basis-design and volumetric-memory challenges absent in 2D.\\

    \section{Methods}\label{sec:method}

KIR-NO reconstructs 3D MRI volumes from undersampled multi-coil $k$-space by combining physics-based data consistency with neural-operator priors. It comprises a forward operator $\mathcal{A}$, an unrolled architecture of data-consistency and image-space neural-operator updates, and an optional $k$-space neural-operator head. To transfer across discretizations while preserving local structure, the neural operators are parameterized as local integration operators via 3D DISCO convolutions. 

We denote the MRI forward imaging operator by $\mathcal{A}$ and model the acquired subsampled $k$-space measurements as $\tilde{\K} = \mathcal{A}(\x_\star) + \epsilon$, where $\x_\star$ is the unknown ground-truth image and $\epsilon$ represents measurement noise and modeling error. Here, $\mathcal{A}$ consists of the multi-coil sensitivity encoding, followed by a Fourier transform and the application of an undersampling mask. Classical compressed sensing MRI reconstructs an image by solving a regularized inverse problem $\hat{\x} = \argmin_{\x} f(\x) + g(\x)$, where the data-fidelity term $f(\x)=\tfrac{1}{2}\big\|\mathcal{A}(\x) - \tilde{\K}\big\|_2^2$ enforces agreement with the acquired measurements, while $g(\x)=\lambda \Psi(\x)$ encodes prior knowledge about plausible images, such as sparsity~\citep{chan2006total}.

\subsection{Unrolled Neural-Operator Reconstruction}

Motivated by this optimization perspective, KIR-NO (Fig.~\ref{fig:arch}) uses an unrolled neural operator: instead of a hand-crafted regularizer $\Psi$, it learns the prior from data while keeping explicit dependence on the acquisition model through data-consistency updates.

\textbf{$k$-space Neural Operator.} We first augment KIR-NO with a $k$-space neural-operator (kNO) head that operates directly on the coil-combined undersampled measurements before any inverse Fourier transform. Given the acquired subsampled $k$-space measurements $\tilde{\K}$, we compute
\begin{align}\label{eq:kno_head}
    \tilde{\K}' = \tilde{\K} + \mathrm{NO}_{k}(\tilde{\K}),
\end{align}
where $\mathrm{NO}_{k}$ denotes a neural operator acting on $k$-space samples. The refined multi-coil measurements $\tilde{\K}'$ are then mapped to image space by the adjoint operator, $\x^0 = \mathcal{A}^{*}(\tilde{\K}')$ (an inverse Fourier transform followed by adjoint coil combination), providing the initial image estimate that is passed to the subsequent image-space cascades.

Operating directly in $k$-space lets the head exploit acquisition-specific structure (e.g., omitted phase-encoding lines or sampling gaps) and supply a learned Fourier-domain initialization of missing spectral content before image-space refinement. Empirically this helps: KIR-NO outperforms the image-space-only variant (\emph{KIR-NO w/o $k$NO}) by up to $+6.6\,\mathrm{dB}$ at full resolution (Table~\ref{tab:kno}).

\textbf{Image-Space Cascades.} After the kNO head, KIR-NO unrolls a sequence of image-space cascades, each representing one learned reconstruction update. 

Starting from the initial image estimate $\x^0 = \mathcal{A}^{*}(\tilde{\K}')$ produced by the kNO head (or the adjoint $\mathcal{A}^{*}(\tilde{\K})$ when the kNO head is omitted), each cascade updates the current reconstruction according to
\begin{align}\label{eq:unrolled_update}
    \x^{\ell+1}
    =
    \x^\ell
    -
    \eta^\ell
    \mathcal{A}^*
    \big(
        \mathcal{A}(\x^\ell) - \tilde{\K}
    \big)
    +
    \lambda^\ell
    \mathrm{NO}_i^\ell(\x^\ell),
\end{align}
where 
\begin{enumerate}
    \item $\mathcal{A}^*\big(\mathcal{A}(\x^\ell) - \tilde{\K}\big)$ is a data-fidelity term that backprojects the measurement residual into image space via the adjoint of $\mathcal{A}$;
    \item $\mathrm{NO}_i^\ell(\x^\ell)$ is a learned neural-operator prior replacing the hand-crafted regularizer~$\Psi$ (removing aliasing, suppressing noise, recovering missing content);
    \item $\eta^\ell,\lambda^\ell$ weight the fidelity and prior terms, respectively.
\end{enumerate}

The kNO head and image-space cascades are trained jointly end-to-end, forming a single reconstruction model rather than independent optimization iterations with fixed update rules. At inference, the trained KIR-NO maps $k$-space measurements $\tilde{\K}$ to a reconstructed 3D image~$\hat{\x}$.\\

\subsection{Neural Operator Parameterization}

Both the image-space cascades and the kNO head use neural operators. Unlike standard networks, which learn maps between fixed-dimensional arrays tied to the training grid~\citep{jatyani2025nomri}, neural operators learn maps between functions~\citep{azizzadenesheli2024neural,berner2025principled} and can be queried at any discretization --- exactly what our coarse-to-fine (train-low, deploy-high) setting requires.

Because MRI reconstruction also requires preserving local anatomical structure, we use local neural operators based on discrete-continuous convolutions (DISCO). This combines resolution-agnostic operator learning with convolutional locality, and serves as the building block for both the image-space cascades and the $k$-space neural-operator head.\\

\subsubsection{Local Integration Operators}

Local inductive bias is usually imposed with compactly supported convolutional kernels (as in CNNs), but standard discrete kernels are not resolution-agnostic: \citet{local_no} show they converge to pointwise linear operators as resolution increases, rather than to a local integration operator.

For a kernel $\kappa$ and input function $g$ defined over a compact domain $D \subset \mathbb{R}^d$, the \textit{local convolution operator} maps $g$ to an output at location $v$. Given input points $(u_j)_{j=1}^m \subset D$ with quadrature weights $q_j$ and output points $v_i \in D$, DISCO~\citep{ocampo2022, local_no} approximates this operator by
\begin{equation}
    \label{eq:disc_conv}
    (\kappa \star g)(v_i)
    =
    \int_D \kappa(u - v_i) \cdot g(u) \, \mathrm{d}u
    \approx
    \sum_{j=1}^m \kappa(u_j - v_i) \cdot g(u_j) q_j .
\end{equation}
We parameterize the kernel as a linear combination of basis functions $
    \kappa(\mathbf{x}) = \sum_{\ell = 1}^L \theta^{\ell} \kappa^{\ell}(\mathbf{x})$ where the coefficients $\theta^{\ell}$ are learnable parameters and the basis functions $\kappa^{\ell}$ are defined in continuous space. The kernel is therefore parameterized independently of the grid on which it is evaluated. For a given resolution, we sample the continuous kernel on the corresponding 3D grid to obtain a discrete convolution stencil that can be applied with standard GPU-accelerated convolution routines. Across resolutions, only this sampled stencil changes; the learned coefficients $\theta^{\ell}$ remain the same. This provides a resolution-agnostic local integration operator, which we use as the building block for both the measurement-space and image-space components of KIR-NO.\\

\subsubsection{DISCO Kernel Bases} \label{sec:methods_basis}

\begin{figure}[t!] 
    \centering
    \includegraphics[width=\linewidth]{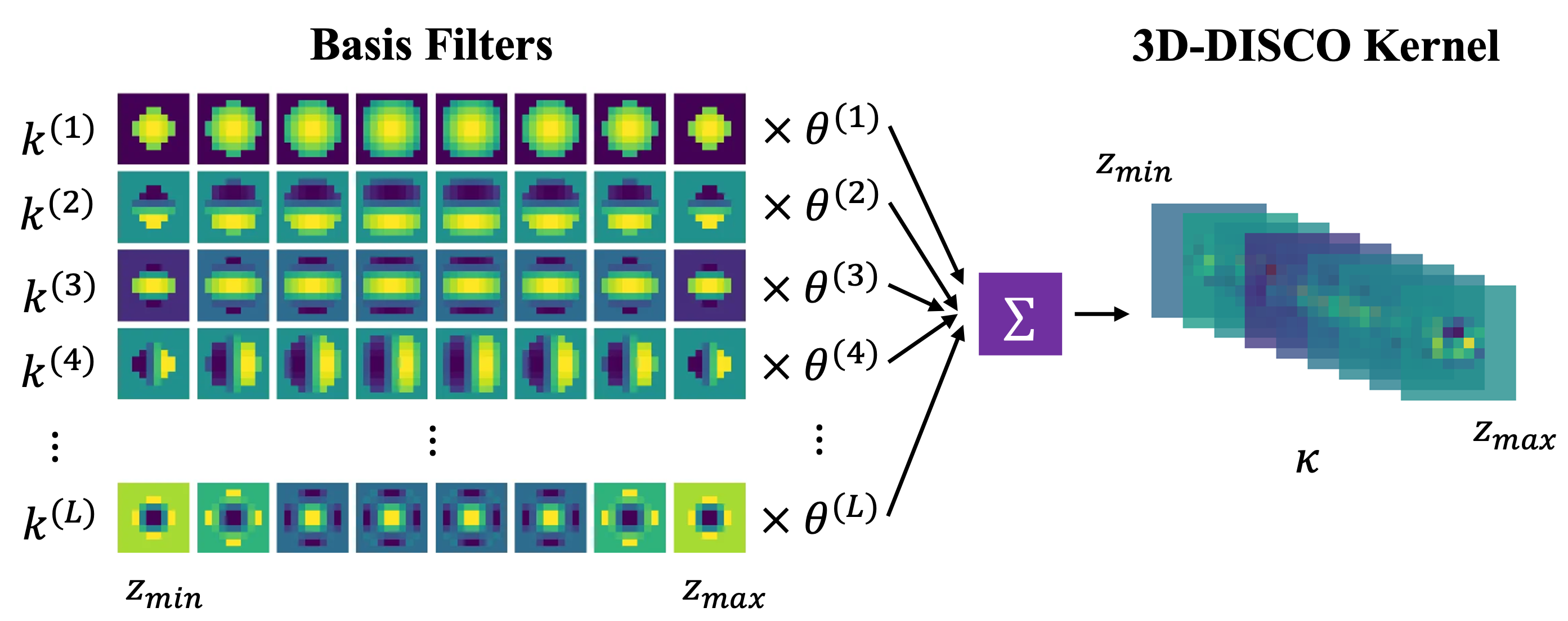}
    \caption{{\bf 3D Morlet-wavelet DISCO basis.} Left: cross-section slices through individual basis functions. Right: a 3D DISCO kernel formed as a learned linear combination of these functions. The discretization shown is for illustration only; the basis is continuous, so the same learned coefficients can be sampled at any target resolution.}
    \label{fig:disco_compile} 
\end{figure}

We introduce two continuous bases of functions used to parameterize the 3D kernels in KIR-NO: 1) a Morlet-wavelet basis, which provides smooth, oriented, and frequency-selective filters. 2) a piecewise linear basis, which provides strictly local interpolation over learned control nodes.\\

\textbf{3D Morlet-Wavelet DISCO Kernels.}
This extends the 2D basis of~\citet{local_no} to 3D. Although referred to as a ``Morlet’’ basis for consistency with~\citet{local_no}, it is more precisely a Hann-windowed real Fourier basis rather than a Gaussian-modulated complex Morlet wavelet. Each basis function is constructed as a localized oscillatory pattern: a separable product of one-dimensional Fourier modes along the $x_1$, $x_2$, and $x_3$ coordinate directions defines the oscillatory component, while a compactly supported Hann-type window localizes the filter in space. Specifically, for local offset $\mathbf{x} = (x_1, x_2, x_3)$ and radial distance $r = \|\mathbf{x}\|_2$, we define
\begin{align}
    \kappa^{p,m,n}_{\mathrm{MW}}(\mathbf{x})
    = W(r)\, H_{p,m,n}(\mathbf{x}),
    \qquad
    W(r) = \mathbf{1}_{[r \leq r_c]} \cos^2\!\left(\pi r/(2 r_c)\right)
\end{align}
with separable Fourier component $H_{p,m,n}(\mathbf{x}) = h_n(x_1)\, h_m(x_2)\, h_p(x_3)$ and 1D factors
\begin{align}
    h_j(s) =
    \cos\!\left(k_j \pi s/w\right) \ \text{for even } j,
    \qquad
    h_j(s) =
    \sin\!\left(k_j \pi s/w\right) \ \text{for odd } j,
\end{align}
where $w$ denotes the half-width of the kernel support along each axis, with $w=r_c$ in our implementation, and $k_j$ is the harmonic order for index $j$. Even indices $j \in \{p,m,n\}$ select cosine modes and odd indices select sine modes, so the multi-index $(p,m,n)$ enumerates combinations of real Fourier modes across the three spatial directions. Larger $k_j$ yield higher-frequency basis functions, while the window $W$ localizes them within the support radius $r_c$. 

The final kernel is learned as a linear combination of these filters and can be discretized at arbitrary resolutions. Figure~\ref{fig:disco_compile} shows representative basis filters and the resulting DISCO kernel construction.\\

\textbf{3D Piecewise Linear DISCO Kernels.}
This basis represents each continuous kernel with local piecewise-linear ``hat'' functions $\Lambda(t) = \max\{1 - |t|, 0\}$ in spherical coordinates centered at the convolution point. For a local offset $\mathbf{x}$, let $(r, \theta, \varphi)$ denote its spherical coordinates, with $r \in [0, r_c]$, $\theta \in [0, \pi]$, and $\varphi \in [-\pi, \pi)$. For kernel shape $(n_r, n_\theta, n_\varphi)$, control nodes lie on $n_r$ evenly spaced radii $r_a = a\,\Delta r$, $a = 0, \dots, n_r{-}1$, with $\Delta r = r_c/(n_r{-}1)$. Since the angles are undefined at $r=0$, the origin carries a single isotropic node, while each of the $n_r{-}1$ outer shells carries an $n_\theta \times n_\varphi$ angular grid
\begin{equation}
    \theta_b = \left(b - \tfrac{1}{2}\right)\Delta \theta, \qquad
    \varphi_c = -\pi + \left(c - \tfrac{1}{2}\right)\Delta \varphi,
\end{equation}
with $\Delta \theta = \pi/n_\theta$ and $\Delta \varphi = 2\pi/n_\varphi$, giving $L = (n_r{-}1)\,n_\theta n_\varphi + 1$ basis functions ($101$ for $[5,5,5]$). 

The origin node defines the isotropic function $\kappa^{0}_{\mathrm{PL}}(\mathbf{x}) = \Lambda(r/\Delta r)$, and each shell node $\ell = (a,b,c)$, $a \geq 1$, defines a product of one-dimensional hat functions with a hard radial cutoff,
\begin{equation}
    \kappa^{\ell}_{\mathrm{PL}}(\mathbf{x})
    = \mathbf{1}_{[r \leq r_c]}\,
      \Lambda\!\left(\tfrac{r - r_a}{\Delta r}\right)
      \Lambda\!\left(\tfrac{\theta - \theta_b}{\Delta \theta}\right)
      \Lambda\!\left(\tfrac{d_{\mathrm{circ}}(\varphi, \varphi_c)}{\Delta \varphi}\right),
\end{equation}
where $d_{\mathrm{circ}}(\varphi, \varphi_c) = \min(|\varphi - \varphi_c|,\, 2\pi - |\varphi - \varphi_c|)$ enforces $2\pi$-periodicity in azimuth.

Rather than mixing fixed oscillatory filters, the model learns nodal kernel values $\theta^{\ell}$ and evaluates the kernel by piecewise multilinear interpolation between nearby control nodes, enabling independent radial and angular resolution. Although less frequency-selective than Morlet, this strictly local basis better matches locally piecewise-smooth MRI images and sparse $k$-space residuals. Used in both the image-space cascades and the $k$-space neural-operator head, it improves over Morlet by $+2.11\,\mathrm{dB}$ PSNR at quarter resolution under an identical kernel configuration and training budget --- despite using fewer coefficients --- so KIR-NO adopts it as the default DISCO parameterization (Table~\ref{tab:basis_ablation}).\\

    \section{Experiments}\label{sec:experiments}

We evaluate on SKM-TEA~\citep{desai2022skm}, a large-scale multi-coil qDESS knee MRI dataset, and BraTS~2021~\citep{brats2021}. Models are trained on quarter- or half-resolution volumes and evaluated zero-shot at the other resolutions, without fine-tuning unless noted. Dataset and preprocessing details appear in Appendix~\ref{app:dataset}; implementation details in Appendix~\ref{app:model}. We compare \emph{KIR-NO} against \emph{(a)} classical reconstruction methods, including Zero-Filled (adjoint SENSE~\citep{SENSE1999}) and $\ell_1$-Wavelet compressed sensing~\citep{lustig2007sparse}; \emph{(b)} conventional CNN baselines, including 2D E2E-VarNet (per-slice), E2E-VarNet3D~\citep{e2evarnet}, and a 3D U-Net~\citep{unet}; and \emph{(c)} diffusion baselines, Diffusion-3D Physics~\citep{krainovic2024resolution} and Diffusion-CSGM~\citep{jalal2021csgm}. For all methods, the coil sensitivity maps are supplied as a known forward operator rather than estimated by the network. They are estimated from scanner calibration for SKM-TEA and simulated for BraTS (Appendix~\ref{app:dataset}). In particular, our E2E-VarNet baselines use these provided maps in place of the original method's learned sensitivity-map module~\citep{e2evarnet}, ensuring that every method shares an identical forward model.\\

\begin{figure*}[t!]

    \centering

    \includegraphics[width=\linewidth]{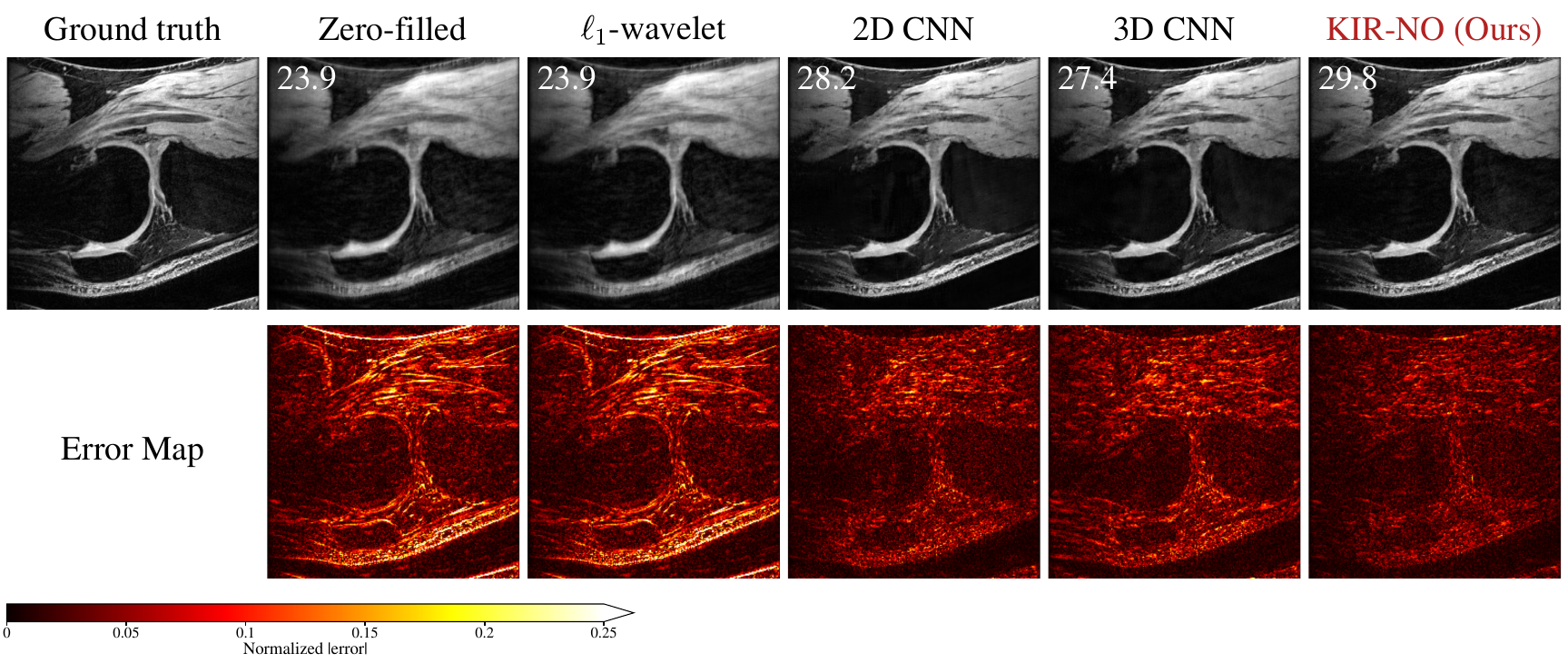}\\[2mm]

    \includegraphics[width=\linewidth]{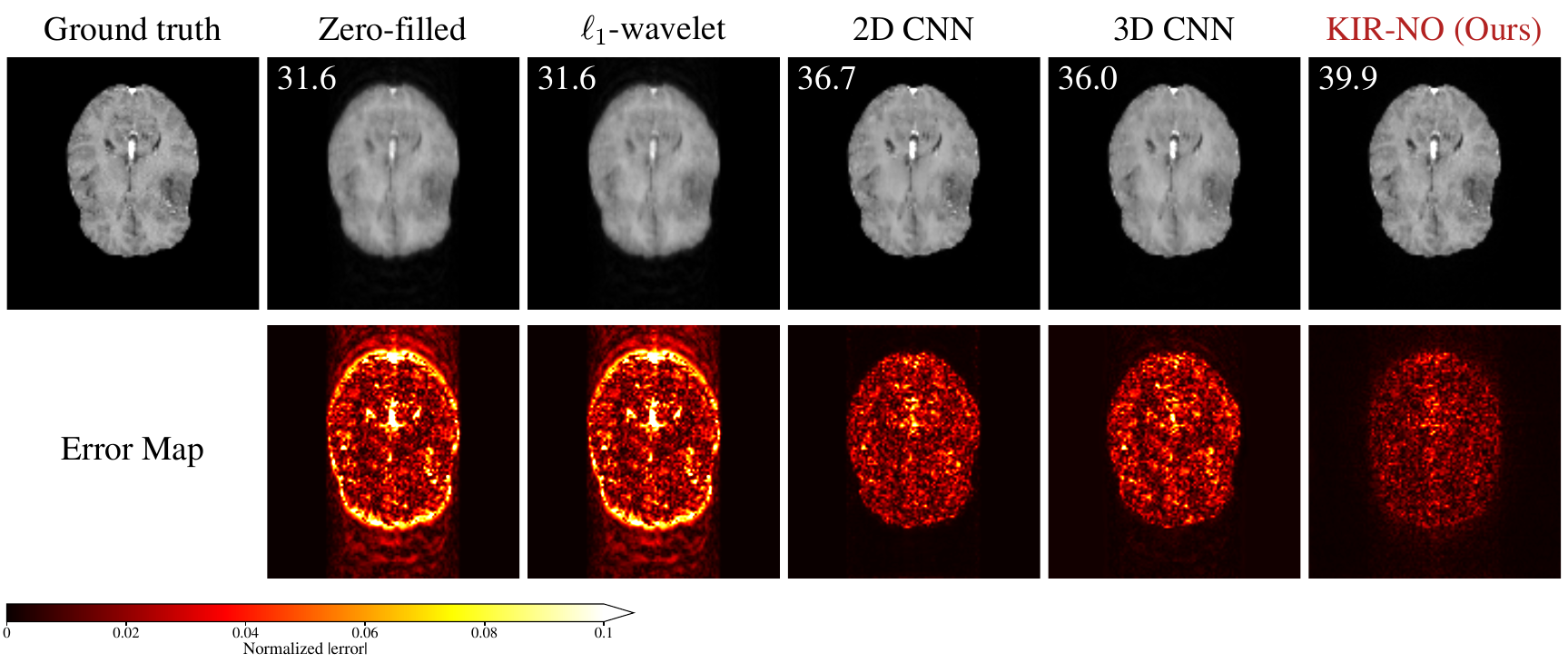}

    \caption{\textbf{Qualitative comparison on knee (top) and brain (bottom) MRI reconstruction.} Reconstructions of a representative slice with corresponding error maps (warmer colors indicate larger residuals). All models are trained and evaluated on half resolution data. The top left number is the PSNR (dB) on the shown slice. The ground truth (GT) is compared to classical baselines (Zero-Filled, $\ell_1$-Wavelet) and learned baselines --- 2D E2E-VarNet, 3D E2E-VarNet, and our KIR-NO. Classical methods show strong aliasing; KIR-NO has the smallest residuals on both samples. Dataset-level results are in Tables~\ref{tab:zero_shot_sr} and~\ref{tab:brats}.}
\label{fig:qualitative}

\end{figure*}

\textbf{Two axes of discretization-agnosticism.} Our experiments isolate two independent axes. \emph{(i) Image-space resolution} (Tables~\ref{tab:zero_shot_sr},~\ref{tab:brats}): the image-space operators ($\mathrm{NO}_i$) are re-sampled on a finer voxel grid at inference, so a model trained at $256$ reconstructs at $512$ with no retraining (the zero-shot super-resolution setting). Crucially, at every evaluation resolution, the k-space head its not re-discretized: it keeps its trained kernel at ever evaluation resolution and only the image head after the inverse transform is resampled onto the finer grid. The model thus receives no additional measurements at higher resolution, so any gain at full resolution comes from the learned model rather than from additional acquired frequencies. Because the image-space kernels are continuous, they are sampled at the new voxel spacing and keep the same physical support, whereas a CNN applies its fixed discrete kernels on the finer grid, which shrinks their receptive field in physical units. \emph{(ii) $k$-space sampling rate} (Table~\ref{tab:accel_rates}): at fixed image resolution ($256$), one trained model handles all undersampling factors ($4\times/8\times/16\times$) zero-shot, only the mask changing, exercising the $k$-space head ($\mathrm{NO}_k$) and data-consistency at corresponding sampling rate. In neither case is $k$-space interpolated to change resolution; sub-native resolutions are formed by image-space resizing followed by an FFT (Appendix~\ref{app:dataset}).\\

\textbf{Zero-shot super-resolution on SKM-TEA.} Table~\ref{tab:zero_shot_sr} reports both training regimes. Trained at half resolution, KIR-NO leads decisively at the in-domain half setting ($36.54$\,dB / $0.924$ SSIM; $+2.23$\,dB over E2E-VarNet3D and $+1.50$\,dB over the strongest CNN) and, most notably, under zero-shot \emph{upward} super-resolution to full resolution ($38.36$\,dB / $0.940$ SSIM, best on all three metrics, $+3.03$\,dB over the best CNN). The CNN baselines transfer stably in PSNR at full resolution (all $\sim$$35.2$--$35.3$\,dB) but lose structural fidelity (3D U-Net SSIM drops to $0.827$), whereas KIR-NO's SSIM rises to $0.940$. At the coarsest zero-shot-\emph{downward} setting (quarter), KIR-NO is competitive ($29.64$\,dB) though the CNNs are marginally higher on this grid. Trained at quarter resolution, KIR-NO again leads at in-domain quarter ($33.13$\,dB / $0.894$) and at full-resolution SR ($35.52$\,dB / $0.908$).\\

\begin{table}[t]
  \caption{{Zero-shot super-resolution on SKM-TEA (3D Knee MRI), $4\times$ acceleration.}
KIR-NO is trained \emph{once per regime} and evaluated at all three resolutions \emph{without retraining or fine-tuning}; the in-domain (training) resolution is indicated in each panel. Classical methods are training-free. Table~\ref{tab:recent_baselines} extends the comparison. Single-cascade, single-step models are reported here; deeper-cascade and multi-step results appear in Appendices~\ref{app:deepcas} and~\ref{app:diffusion}.}
    \label{tab:zero_shot_sr}
    \centering
    \footnotesize
    \resizebox{\linewidth}{!}{
        \begin{tabular}{l ccc ccc ccc}
            \toprule
            & \multicolumn{3}{c}{\textbf{Quarter Res (128)}}
            & \multicolumn{3}{c}{\textbf{Half Res (256)}}
            & \multicolumn{3}{c}{\textbf{Full Res SR (512)}} \\
            \cmidrule(lr){2-4} \cmidrule(lr){5-7} \cmidrule(lr){8-10}
            \textbf{Method} & PSNR$\uparrow$ & SSIM$\uparrow$ & NMSE$\downarrow$
                            & PSNR$\uparrow$ & SSIM$\uparrow$ & NMSE$\downarrow$
                            & PSNR$\uparrow$ & SSIM$\uparrow$ & NMSE$\downarrow$ \\
            \midrule
            \multicolumn{10}{l}{\textit{Learning-Free Methods (training-free)}} \\
            \cmidrule(lr){1-10}
            Zero-Filled      & $26.59_{\pm 1.31}$ & $0.700_{\pm 0.036}$ & $0.1074_{\pm 0.0222}$
                             & $29.80_{\pm 1.49}$ & $0.797_{\pm 0.031}$ & $0.0616_{\pm 0.0148}$
                             & $34.46_{\pm 1.30}$ & $0.882_{\pm 0.016}$ & $0.0241_{\pm 0.0061}$ \\
            $\ell_1$-Wavelet & $26.62_{\pm 1.31}$ & $0.701_{\pm 0.036}$ & $0.1067_{\pm 0.0221}$
                             & $29.82_{\pm 1.49}$ & $0.797_{\pm 0.031}$ & $0.0612_{\pm 0.0147}$
                             & $34.47_{\pm 1.30}$ & $0.882_{\pm 0.016}$ & $0.0240_{\pm 0.0061}$ \\
            \midrule
            \multicolumn{10}{l}{\textit{Trained @ half resolution \quad (in-domain: Half;\ \ zero-shot: Quarter, Full)}} \\
            \cmidrule(lr){1-10}
            2D E2E-VarNet (CNN)    & $\mathbf{30.41}_{\pm 1.47}$ & $\mathbf{0.849}_{\pm 0.029}$ & $\mathbf{0.0452}_{\pm 0.0122}$
                                   & $35.04_{\pm 1.58}$ & $0.906_{\pm 0.020}$ & $0.0186_{\pm 0.0054}$
                                   & $35.33_{\pm 1.24}$ & $0.896_{\pm 0.016}$ & $0.0195_{\pm 0.0041}$ \\
            3D E2E-VarNet (CNN)    & $30.08_{\pm 1.55}$ & $0.828_{\pm 0.040}$ & $0.0490_{\pm 0.0140}$
                                   & $34.31_{\pm 1.73}$ & $0.883_{\pm 0.027}$ & $0.0223_{\pm 0.0073}$
                                   & $35.32_{\pm 1.38}$ & $0.870_{\pm 0.025}$ & $0.0199_{\pm 0.0060}$ \\
            3D U-Net               & $30.33_{\pm 1.39}$ & $0.847_{\pm 0.026}$ & $0.0458_{\pm 0.0114}$
                                   & $34.64_{\pm 1.51}$ & $0.879_{\pm 0.017}$ & $0.0204_{\pm 0.0056}$
                                   & $35.24_{\pm 1.23}$ & $0.827_{\pm 0.015}$ & $0.0200_{\pm 0.0048}$ \\
            Diffusion-3D Physics & $26.99_{\pm 1.33}$ & $0.716_{\pm 0.035}$ & $0.0982_{\pm 0.0214}$
                                   & $32.42_{\pm 1.57}$ & $0.827_{\pm 0.027}$ & $0.0341_{\pm 0.0104}$
                                   & $35.02_{\pm 1.30}$ & $0.892_{\pm 0.015}$ & $0.0212_{\pm 0.0056}$ \\
            Diffusion-CSGM & $26.61_{\pm 1.31}$ & $0.701_{\pm 0.036}$ & $0.1069_{\pm 0.0222}$
                                   & $29.89_{\pm 1.50}$ & $0.799_{\pm 0.031}$ & $0.0603_{\pm 0.0146}$
                                   & $34.61_{\pm 1.31}$ & $0.885_{\pm 0.016}$ & $0.0233_{\pm 0.0060}$ \\
            \textbf{KIR-NO (Ours)} & $29.64_{\pm 1.27}$ & $0.770_{\pm 0.027}$ & $0.0534_{\pm 0.0115}$
                                   & $\mathbf{36.54}_{\pm 1.60}$ & $\mathbf{0.924}_{\pm 0.016}$ & $\mathbf{0.0132}_{\pm 0.0040}$
                                   & $\mathbf{38.36}_{\pm 1.25}$ & $\mathbf{0.940}_{\pm 0.010}$ & $\mathbf{0.0097}_{\pm 0.0021}$ \\
            \midrule
            \multicolumn{10}{l}{\textit{Trained @ quarter resolution \quad (in-domain: Quarter;\ \ zero-shot: Half, Full)}} \\
            \cmidrule(lr){1-10}
            2D E2E-VarNet (CNN)    & $31.45_{\pm 1.46}$          & $0.874_{\pm 0.027}$          & $0.0357_{\pm 0.0099}$
                                   & $33.29_{\pm 1.43}$          & $\mathbf{0.878}_{\pm 0.023}$          & $0.0275_{\pm 0.0068}$
                                   & $33.11_{\pm 1.19}$          & $0.852_{\pm 0.020}$          & $0.0325_{\pm 0.0063}$ \\
            3D E2E-VarNet (CNN)    & $30.65_{\pm 1.50}$          & $0.852_{\pm 0.030}$          & $0.0430_{\pm 0.0124}$
                                   & $33.11_{\pm 1.52}$          & $0.850_{\pm 0.040}$          & $0.0291_{\pm 0.0088}$
                                   & $34.35_{\pm 1.40}$          & $0.845_{\pm 0.040}$          & $0.0252_{\pm 0.0088}$ \\
            3D U-Net               & $31.43_{\pm 1.43}$          & $0.860_{\pm 0.022}$          & $0.0358_{\pm 0.0100}$
                                   & $33.63_{\pm 1.46}$          & $0.859_{\pm 0.023}$          & $0.0256_{\pm 0.0069}$
                                   & $34.88_{\pm 1.28}$          & $0.855_{\pm 0.027}$          & $0.0218_{\pm 0.0052}$ \\
            \textbf{KIR-NO (Ours)} & $\mathbf{33.13}_{\pm 1.44}$ & $\mathbf{0.894}_{\pm 0.020}$ & $\mathbf{0.0242}_{\pm 0.0069}$
                                   & $\mathbf{33.69}_{\pm 1.45}$ & $0.872_{\pm 0.021}$ & $\mathbf{0.0251}_{\pm 0.0060}$
                                   & $\mathbf{35.52}_{\pm 1.22}$ & $\mathbf{0.908}_{\pm 0.014}$ & $\mathbf{0.0187}_{\pm 0.0038}$ \\
            \bottomrule
        \end{tabular}} \vspace{3mm}
\end{table}

\textbf{Zero-shot super-resolution on BraTS.} Table~\ref{tab:brats} applies the same two-regime protocol to BraTS brain MRI ($8$-coil simulated $k$-space); every row is evaluated on the same $20$ test volumes at all three resolutions, and the E2E-VarNets are trained with the same budget as KIR-NO ($100$ volumes per epoch for $120$ epochs). Trained at quarter resolution, KIR-NO leads in-domain by $+1.67$\,dB over the 2D E2E-VarNet ($40.14$ vs $38.47$\,dB) with the best NMSE ($0.0075$), while the CNN baselines transfer upward more strongly on this simulated $8$-coil data (2D E2E-VarNet $39.84$ / $41.42$\,dB vs KIR-NO $39.07$ / $40.34$\,dB at half / full resolution). Trained at half resolution, KIR-NO leads in-domain ($43.37$\,dB; $+1.24$\,dB over the 2D and $+2.41$\,dB over the 3D E2E-VarNet) and at full resolution ($43.72$\,dB; $+0.05$ / $+0.64$\,dB), with the best NMSE at both. Figure~\ref{fig:qualitative} shows representative reconstructions and error maps for both anatomies.\\

\begin{table}[t]
    \centering
    \caption{Zero-shot super-resolution on BraTS~2021 (3D Brain MRI) reconstruction at $4\times$ acceleration, {8-coil}. KIR-NO is trained \emph{once per regime} and evaluated at all three resolutions; the in-domain (training) resolution is indicated in each panel.}
    \label{tab:brats}
    \resizebox{\linewidth}{!}{
    \begin{tabular}{l ccc ccc ccc}
    \toprule
    & \multicolumn{3}{c}{\textbf{Quarter (60)}}
    & \multicolumn{3}{c}{\textbf{Half (120)}}
    & \multicolumn{3}{c}{\textbf{Full (240)}} \\
    \cmidrule(lr){2-4}\cmidrule(lr){5-7}\cmidrule(lr){8-10}
    \textbf{Method}
    & PSNR$\uparrow$ & SSIM$\uparrow$ & NMSE$\downarrow$
    & PSNR$\uparrow$ & SSIM$\uparrow$ & NMSE$\downarrow$
    & PSNR$\uparrow$ & SSIM$\uparrow$ & NMSE$\downarrow$ \\
    \midrule
    \multicolumn{10}{l}{\textit{Learning-Free Methods (training-free)}} \\
        \midrule
        Zero-Filled            & $30.59_{\pm 1.45}$ & $0.756_{\pm 0.028}$ & $0.0663_{\pm 0.0135}$ & $34.14_{\pm 1.08}$ & $0.801_{\pm 0.024}$ & $0.0355_{\pm 0.0109}$ & $37.04_{\pm 1.15}$ & $0.853_{\pm 0.022}$ & $0.0243_{\pm 0.0078}$ \\
    $\ell_1$-Wavelet & $30.60_{\pm 1.45}$ & $0.756_{\pm 0.028}$ & $0.0661_{\pm 0.0134}$ & $34.16_{\pm 1.08}$ & $0.800_{\pm 0.024}$ & $0.0354_{\pm 0.0108}$ & $37.05_{\pm 1.15}$ & $0.852_{\pm 0.022}$ & $0.0242_{\pm 0.0078}$ \\
    \midrule
    \multicolumn{10}{l}{\textit{Trained @ half resolution \quad (in-domain: Half;\ \ zero-shot: Quarter, Full)}} \\ 
    \cmidrule(lr){1-10}

    2D E2E-VarNet (CNN)    & $36.51_{\pm 1.65}$ & $\mathbf{0.939}_{\pm 0.014}$ & $0.0170_{\pm 0.0038}$ & $42.13_{\pm 1.28}$ & $\mathbf{0.981}_{\pm 0.004}$ & $0.0057_{\pm 0.0017}$ & $43.67_{\pm 1.43}$ & $\mathbf{0.974}_{\pm 0.007}$ & $0.0053_{\pm 0.0017}$ \\
    3D E2E-VarNet (CNN)    & $\mathbf{36.70}_{\pm 1.51}$ & $0.936_{\pm 0.012}$ & $\mathbf{0.0163}_{\pm 0.0039}$ & $40.96_{\pm 1.25}$ & $0.960_{\pm 0.007}$ & $0.0074_{\pm 0.0022}$ & $43.08_{\pm 1.20}$ & $0.973_{\pm 0.006}$ & $0.0059_{\pm 0.0014}$ \\
    Diffusion-3D Physics   & $30.80_{\pm 1.43}$ & $0.674_{\pm 0.055}$ & $0.0631_{\pm 0.0130}$ & $34.50_{\pm 1.08}$ & $0.823_{\pm 0.026}$ & $0.0328_{\pm 0.0102}$ & $37.28_{\pm 1.16}$ & $0.881_{\pm 0.020}$ & $0.0230_{\pm 0.0074}$ \\
    Diffusion-CSGM         & $30.68_{\pm 1.45}$ & $0.759_{\pm 0.028}$ & $0.0649_{\pm 0.0133}$ & $34.28_{\pm 1.08}$ & $0.803_{\pm 0.024}$ & $0.0345_{\pm 0.0107}$ & $37.21_{\pm 1.16}$ & $0.855_{\pm 0.022}$ & $0.0234_{\pm 0.0077}$ \\
    \textbf{KIR-NO (Ours)} & $34.50_{\pm 1.76}$ & $0.828_{\pm 0.040}$ & $0.0265_{\pm 0.0028}$ & $\mathbf{43.37}_{\pm 1.34}$ & $0.960_{\pm 0.007}$ & $\mathbf{0.0043}_{\pm 0.0016}$ & $\mathbf{43.72}_{\pm 1.12}$ & $0.933_{\pm 0.011}$ & $\mathbf{0.0052}_{\pm 0.0015}$ \\
    \midrule
    \multicolumn{10}{l}{\textit{Trained @ quarter resolution \quad (in-domain: Quarter;\ \ zero-shot: Half, Full)}} \\
    \cmidrule(lr){1-10}

    2D E2E-VarNet (CNN)    & $38.47_{\pm 1.55}$ & $\mathbf{0.969}_{\pm 0.004}$ & $0.0109_{\pm 0.0026}$ & $\mathbf{39.84}_{\pm 1.26}$ & $0.953_{\pm 0.014}$ & $\mathbf{0.0095}_{\pm 0.0027}$ & $\mathbf{41.42}_{\pm 1.19}$ & $0.958_{\pm 0.013}$ & $\mathbf{0.0086}_{\pm 0.0022}$ \\
    3D E2E-VarNet (CNN)    & $37.73_{\pm 1.47}$ & $0.964_{\pm 0.004}$ & $0.0129_{\pm 0.0031}$ & $37.84_{\pm 1.46}$ & $\mathbf{0.972}_{\pm 0.004}$ & $0.0146_{\pm 0.0022}$ & $40.55_{\pm 1.15}$ & $\mathbf{0.979}_{\pm 0.004}$ & $0.0104_{\pm 0.0017}$ \\
    \textbf{KIR-NO (Ours)} & $\mathbf{40.14}_{\pm 1.48}$ & $0.958_{\pm 0.008}$ & $\mathbf{0.0075}_{\pm 0.0022}$ & $39.07_{\pm 1.18}$ & $0.870_{\pm 0.021}$ & $0.0114_{\pm 0.0034}$ & $40.34_{\pm 1.00}$ & $0.864_{\pm 0.019}$ & $0.0111_{\pm 0.0027}$ \\
    \bottomrule
    \end{tabular}} \vspace{3mm}
  \end{table}



\textbf{Robustness to acceleration.}
Table~\ref{tab:accel_rates} reports half-resolution reconstruction at $4\times$, $8\times$, and $16\times$ acceleration. KIR-NO leads at every rate, improving over the 3D CNN baseline by $+2.23\,\mathrm{dB}$, $+2.40\,\mathrm{dB}$, and $+1.97\,\mathrm{dB}$, respectively. KIR-NO retains the highest SSIM at every rate ($0.924$, $0.895$, $0.776$ at $4\times$/$8\times$/$16\times$, versus E2E-VarNet3D's $0.883$, $0.847$, $0.716$), remaining roughly $+0.04$--$0.06$ SSIM ahead across rates.\\

\begin{table}[t!]
    \centering
    \caption{Half-resolution reconstruction at increasing acceleration
             rates on SKM-TEA. Each learned method is represented by a single trained model, evaluated at all three rates without retraining; only the $k$-space sampling mask changes between columns.}
    \label{tab:accel_rates}
    \footnotesize
    \resizebox{\linewidth}{!}{
        \begin{tabular}{lccccccccc}
            \toprule
            & \multicolumn{3}{c}{\textbf{$4\times$}}
            & \multicolumn{3}{c}{\textbf{$8\times$}}
            & \multicolumn{3}{c}{\textbf{$16\times$}} \\
            \cmidrule(lr){2-4} \cmidrule(lr){5-7} \cmidrule(lr){8-10}
            \textbf{Method} & PSNR & SSIM & NMSE & PSNR & SSIM & NMSE & PSNR & SSIM & NMSE \\
            \midrule
            Zero-Filled            & $29.80_{\pm 1.49}$          & $0.797_{\pm 0.031}$          & $0.0616_{\pm 0.0148}$
                                   & $28.74_{\pm 1.47}$          & $0.758_{\pm 0.034}$          & $0.0784_{\pm 0.0178}$
                                   & $26.31_{\pm 1.45}$          & $0.668_{\pm 0.043}$          & $0.1366_{\pm 0.0278}$ \\
            $\ell_1$-Wavelet       & $29.82_{\pm 1.49}$          & $0.797_{\pm 0.031}$          & $0.0612_{\pm 0.0147}$
                                   & $28.76_{\pm 1.47}$          & $0.758_{\pm 0.034}$          & $0.0780_{\pm 0.0177}$
                                   & $26.33_{\pm 1.45}$          & $0.668_{\pm 0.043}$          & $0.1360_{\pm 0.0277}$ \\
            2D E2E-VarNet (CNN) & $35.04_{\pm 1.58}$          & $0.906_{\pm 0.020}$          & $0.0186_{\pm 0.0054}$
                                   & $32.82_{\pm 1.49}$          & $0.873_{\pm 0.024}$          & $0.0307_{\pm 0.0076}$
                                   & $28.16_{\pm 1.45}$          & $0.757_{\pm 0.035}$          & $0.0893_{\pm 0.0188}$ \\
            3D E2E-VarNet (CNN)    & $34.31_{\pm 1.73}$          & $0.883_{\pm 0.027}$          & $0.0223_{\pm 0.0073}$
                                   & $32.14_{\pm 1.66}$          & $0.847_{\pm 0.032}$          & $0.0364_{\pm 0.0105}$
                                   & $27.25_{\pm 1.53}$          & $0.716_{\pm 0.047}$          & $0.1102_{\pm 0.0241}$ \\
            \textbf{KIR-NO (Ours)} & $\mathbf{36.54}_{\pm 1.60}$ & $\mathbf{0.924}_{\pm 0.016}$ & $\mathbf{0.0132}_{\pm 0.0040}$
                                   & $\mathbf{34.54}_{\pm 1.54}$ & $\mathbf{0.895}_{\pm 0.020}$ & $\mathbf{0.0209}_{\pm 0.0058}$
                                   & $\mathbf{29.22}_{\pm 1.47}$ & $\mathbf{0.776}_{\pm 0.035}$ & $\mathbf{0.0703}_{\pm 0.0165}$ \\
            \bottomrule
        \end{tabular}}  \vspace{6mm}
\end{table}

\textbf{Robustness to sampling pattern.}
Table~\ref{tab:pattern_ablation} examines sensitivity to the $k$-space sampling pattern. Beyond the Gaussian-density Cartesian mask used elsewhere, we retrain both E2E-VarNets and KIR-NO for each of equispaced, radial, and 2D Poisson-disc patterns at quarter resolution and $4\times$ acceleration (Appendix~\ref{app:patterns}). KIR-NO attains the highest PSNR and SSIM under every pattern (Table~\ref{tab:pattern_ablation}); the Poisson-disc pattern is the most favorable, reaching $35.30\,\mathrm{dB}$ ($+2.17\,\mathrm{dB}$ over the Gaussian default). This shows that KIR-NO's gains are not specific to the Gaussian Cartesian mask.\\


\textbf{Convergence efficiency.}
Figure~\ref{fig:intro}(e) reports the first epoch at which each model reaches a target PSNR on half-resolution validation. KIR-NO reaches $34\,\mathrm{dB}$ at epoch $5$ versus $31$ for the E2E-VarNet3D CNN baseline, about $6\times$ faster, and reaches $36\,\mathrm{dB}$ at epoch $39$, a threshold the CNN baseline never reaches within $120$ epochs. Combined with the $+2.23\,\mathrm{dB}$ final-PSNR gain at half resolution (Table~\ref{tab:accel_rates}), these results show improved reconstruction quality and convergence efficiency.

\textbf{Resolution-wise PSNR.}
Figure~\ref{fig:intro}(d) plots PSNR at quarter, half, and full resolution for the zero-filled baseline and the quarter-trained 2D CNN, 3D CNN and KIR-NO of Table~\ref{tab:zero_shot_sr}. KIR-NO leads at every resolution ($33.13$, $33.69$ and $35.52$\,dB). At full resolution the 2D and 3D CNN baselines (E2E-VarNet) fall below the training-free zero-filled reconstruction ($33.11$ and $34.35$ vs $34.46$\,dB), while KIR-NO stays $1.06$\,dB above it. The $k$-space head ablation is in Appendix~\ref{app:kno} (Table~\ref{tab:kno}).\\ 
    \section{Conclusion}\label{sec:conclusion}

We introduced \emph{KIR-NO}, a resolution-agnostic neural-operator framework for accelerated 3D MRI reconstruction built on discrete-continuous convolutions (DISCO). By parameterizing kernels in continuous space while preserving locality, KIR-NO enables a coarse-to-fine paradigm: train with full backpropagation on memory-feasible low-resolution volumes, and deploy zero-shot at high resolution. On SKM-TEA, KIR-NO improves reconstruction quality across acceleration rates and under zero-shot super-resolution, and converges faster than the CNN baseline; on BraTS it leads in-domain in both training regimes, and transfers upward more strongly from half resolution training. Ablations show that the $k$-space head drives the high-resolution gains, that additional image-space cascades further improve accuracy, and that the piecewise-linear DISCO basis outperforms the Morlet-wavelet basis.\\

\textbf{Limitations.}
The full 3D model needs large GPU memory, and we use a single image-space cascade by default; gradient checkpointing enables deeper unrolling, which yields further gains for both KIR-NO and the CNN baseline (Table~\ref{tab:cascades_cnn_kno}) at additional compute cost. 
Zero-shot super-resolution operates entirely in image space: the $k$-space head and data-consistency use the fixed training-resolution measurements, and only the post-transform image space operators are resampled onto the finer grid. On SKM-TEA the full-resolution in-plane ground truth contains acquired frequencies well beyond the training extent ($416\times512$ acquisition matrix, zero-padded to $512\times512$ by the dataset), so it is a non-circular target for this prediction; the through-plane axis is interpolated and may inflate metrics there. The evaluation does not capture real-world noise, motion, or coil-geometry shifts of genuinely distinct high-resolution acquisitions. Experiments are restricted to two anatomies (SKM-TEA, BraTS) with single-seed runs; while we ablate different sampling patterns (Table~\ref{tab:pattern_ablation}), multi-contrast acquisitions, more aggressive accelerations, and matched comparisons against the broader score-based and unrolled-network landscape would further strengthen the empirical positioning. \\

\section*{Acknowledgment}
This work is supported in part by ONR (MURI grant N000142312654 and N000142012786). R.D. is supported in part by the Undergraduate Research Fellowships (SURF) at Caltech.  A.A. is supported in part by Bren endowed chair and the AI2050 senior fellow program at Schmidt Sciences.\\
    


\bibliographystyle{unsrtnat}
\bibliography{main.bib}

@article{brats2021,
  title={The {RSNA-ASNR-MICCAI} {BraTS} 2021 Benchmark on Brain Tumor Segmentation and Radiogenomic Classification},
  author={Baid, Ujjwal and Ghodasara, Satyam and Mohan, Suyash and Bilello, Michel and Calabrese, Evan and Colak, Errol and Farahani, Keyvan and Kalpathy-Cramer, Jayashree and Kitamura, Felipe C and Pati, Sarthak and others},
  journal={arXiv preprint arXiv:2107.02314},
  year={2021}
}

@article{datta2025resolution,
  title={Resolution-Independent Neural Operators for Multi-Rate Sparse-View {CT}},
  author={Datta, Aujasvit and Wang, Jiayun and Aali, Asad and Jatyani, Armeet Singh and Anandkumar, Anima},
  journal={arXiv preprint arXiv:2512.12236},
  year={2025}
}

@article{gan2023block,
  title={Block coordinate plug-and-play methods for blind inverse problems},
  author={Gan, Weijie and Shoushtari, Shirin and Hu, Yuyang and Liu, Jiaming and An, Hongyu and Kamilov, Ulugbek S.},
  journal={Advances in Neural Information Processing Systems},
  volume={36},
  pages={78607--78620},
  year={2023}
}

@article{krainovic2024resolution,
  title={Resolution-Robust {3D} {MRI} Reconstruction with {2D} Diffusion Priors: Diverse-Resolution Training Outperforms Interpolation},
  author={Krainovic, Anselm and Ruschke, Stefan and Heckel, Reinhard},
  journal={arXiv preprint arXiv:2412.18584},
  year={2024}
}

@article{jalal2021csgm,
  title={Robust Compressed Sensing {MRI} with Deep Generative Priors},
  author={Jalal, Ajil and Arvinte, Marius and Daras, Giannis and Price, Eric and Dimakis, Alexandros G and Tamir, Jonathan I},
  journal={Advances in Neural Information Processing Systems},
  year={2021}
}

@article{kamilov2023plug,
  title={Plug-and-play methods for integrating physical and learned models in computational imaging: Theory, algorithms, and applications},
  author={Kamilov, Ulugbek S and Bouman, Charles A and Buzzard, Gregery T and Wohlberg, Brendt},
  journal={IEEE Signal Processing Magazine},
  volume={40},
  number={1},
  pages={85--97},
  year={2023},
  publisher={IEEE}
}

@article{chan2006total,
  title={Total variation image restoration: Overview and recent developments},
  author={Chan, Tony and Esedoglu, Selim and Park, Frederick and Yip, Andy},
  journal={Handbook of mathematical models in computer vision},
  pages={17--31},
  year={2006},
  publisher={Springer}
}

@inproceedings{jatyani2025nomri,
  author    = {Armeet Singh Jatyani and Jiayun Wang and Aditi Chandrashekar and Zihui Wu and Miguel Liu-Schiaffini and Bahareh Tolooshams and Anima Anandkumar},
  title     = {A Unified Model for Compressed Sensing {MRI} Across Undersampling Patterns},
  booktitle = {Conference on Computer Vision and Pattern Recognition (CVPR) Proceedings},
  abbr      = {CVPR},
  year      = {2025}
}

@misc{wang2025accelerating3dphotoacousticcomputed,
      title={Accelerating {3D} Photoacoustic Computed Tomography with End-to-End Physics-Aware Neural Operators}, 
      author={Jiayun Wang and Yousuf Aborahama and Arya Khokhar and Yang Zhang and Chuwei Wang and Karteekeya Sastry and Julius Berner and Yilin Luo and Boris Bonev and Zongyi Li and Kamyar Azizzadenesheli and Lihong V. Wang and Anima Anandkumar},
      year={2025},
      eprint={2509.09894},
      archivePrefix={arXiv},
      primaryClass={eess.IV},
      url={https://arxiv.org/abs/2509.09894}, 
}

@article{husband2001mri,
	title={{MRI} in the diagnosis and treatment of suspected malignant spinal cord compression},
	author={Husband, DJ and Grant, KA and Romaniuk, CS},
	journal={The British journal of radiology},
	volume={74},
	pages={15--23},
	year={2001},
	publisher={British Institute of Radiology}
}

@article{johnson2019conditional,
	title={Conditional generative adversarial network for {3D} rigid-body motion correction in {MRI}},
	author={Johnson, Patricia M and Drangova, Maria},
	journal={Magnetic resonance in medicine},
	volume={82},
	number={3},
	pages={901--910},
	year={2019},
	publisher={Wiley Online Library}
}

@article{sun2016deep,
	title={Deep {ADMM-Net} for compressive sensing {MRI}},
	author={Yang, Yan and Sun, Jian and Li, Huibin and Xu, Zongben},
	journal={Advances in neural information processing systems},
	volume={29},
	year={2016}
}

@article{lustig2008compressed,
	title={Compressed sensing {MRI}},
	author={Lustig, Michael and Donoho, David L and Santos, Juan M and Pauly, John M},
	journal={IEEE signal processing magazine},
	volume={25},
	number={2},
	pages={72--82},
	year={2008},
	publisher={IEEE}
}

@article{chung2022score,
	title={Score-based diffusion models for accelerated {MRI}},
	author={Chung, Hyungjin and Ye, Jong Chul},
	journal={Medical Image Analysis},
	pages={102479},
	year={2022},
	publisher={Elsevier}
}

@article{griswold2002generalized,
	title={Generalized autocalibrating partially parallel acquisitions ({GRAPPA})},
	author={Griswold, Mark A and Jakob, Peter M and Heidemann, Robin M and Nittka, Mathias and Jellus, Vladimir and Wang, Jianmin and Kiefer, Berthold and Haase, Axel},
	journal={Magnetic Resonance in Medicine: An Official Journal of the International Society for Magnetic Resonance in Medicine},
	volume={47},
	number={6},
	pages={1202--1210},
	year={2002},
	publisher={Wiley Online Library}
}

@article{lustig2007sparse,
	title={Sparse {MRI}: The application of compressed sensing for rapid {MR} imaging},
	author={Lustig, Michael and Donoho, David and Pauly, John M},
	journal={Magnetic Resonance in Medicine: An Official Journal of the International Society for Magnetic Resonance in Medicine},
	volume={58},
	number={6},
	pages={1182--1195},
	year={2007},
	publisher={Wiley Online Library}
}

@article{SENSE1999,
author = {Pruessmann, Klaas P. and Weiger, Markus and Scheidegger, Markus B. and Boesiger, Peter},
title = {{SENSE}: Sensitivity encoding for fast {MRI}},
journal = {Magnetic Resonance in Medicine},
volume = {42},
number = {5},
pages = {952-962},
year = {1999}
}

@article{hammernik2018learning,
	title={Learning a variational network for reconstruction of accelerated {MRI} data},
	author={Hammernik, Kerstin and Klatzer, Teresa and Kobler, Erich and Recht, Michael P and Sodickson, Daniel K and Pock, Thomas and Knoll, Florian},
	journal={Magnetic resonance in medicine},
	volume={79},
	number={6},
	pages={3055--3071},
	year={2018},
	publisher={Wiley Online Library}
}

@inproceedings{e2evarnet,
	title={End-to-end variational networks for accelerated {MRI} reconstruction},
	author={Sriram, Anuroop and Zbontar, Jure and Murrell, Tullie and Defazio, Aaron and Zitnick, C Lawrence and Yakubova, Nafissa and Knoll, Florian and Johnson, Patricia},
	booktitle={Medical Image Computing and Computer Assisted Intervention--MICCAI 2020: 23rd International Conference, Lima, Peru, October 4--8, 2020, Proceedings, Part II 23},
	pages={64--73},
	year={2020},
	organization={Springer}
}

@misc{zbontar2018fastMRI,
	title={{fastMRI}: An Open Dataset and Benchmarks for Accelerated {MRI}},
	author={Jure Zbontar and Florian Knoll and Anuroop Sriram and Tullie Murrell and Zhengnan Huang and Matthew J. Muckley and Aaron Defazio and Ruben Stern and Patricia Johnson and Mary Bruno and Marc Parente and Krzysztof J. Geras and Joe Katsnelson and Hersh Chandarana and Zizhao Zhang and Michal Drozdzal and Adriana Romero and Michael Rabbat and Pascal Vincent and Nafissa Yakubova and James Pinkerton and Duo Wang and Erich Owens and C. Lawrence Zitnick and Michael P. Recht and Daniel K. Sodickson and Yvonne W. Lui},
	journal = {ArXiv e-prints},
	archivePrefix = "arXiv",
	eprint = {1811.08839},
	year={2018}
}

@misc{unet,
	title={{U-Net}: Convolutional Networks for Biomedical Image Segmentation},
	author={Olaf Ronneberger and Philipp Fischer and Thomas Brox},
	year={2015},
	eprint={1505.04597},
	archivePrefix={arXiv},
	primaryClass={cs.CV},
	url={https://arxiv.org/abs/1505.04597},
}

@article{azizzadenesheli2024neural,
	title={Neural operators for accelerating scientific simulations and design},
	author={Azizzadenesheli, Kamyar and Kovachki, Nikola and Li, Zongyi and Liu-Schiaffini, Miguel and Kossaifi, Jean and Anandkumar, Anima},
	journal={Nature Reviews Physics},
	pages={1--9},
	year={2024},
	publisher={Nature Publishing Group UK London}
}

@misc{fnop,
	title={Fourier Neural Operator for Parametric Partial Differential Equations},
	author={Zongyi Li and Nikola Kovachki and Kamyar Azizzadenesheli and Burigede Liu and Kaushik Bhattacharya and Andrew Stuart and Anima Anandkumar},
	year={2021},
	eprint={2010.08895},
	archivePrefix={arXiv},
	primaryClass={cs.LG}
}

@inproceedings{local_no,
	title={Neural Operators with Localized Integral and Differential Kernels},
	author={Liu-Schiaffini, Miguel and Berner, Julius and Bonev, Boris and Kurth, Thorsten and Azizzadenesheli, Kamyar and Anandkumar, Anima},
	booktitle={Forty-first International Conference on Machine Learning},
	year={2024}
}

@article{ocampo2022,
	title={Scalable and equivariant spherical {CNNs} by discrete-continuous ({DISCO}) convolutions},
	author={Ocampo, Jeremy and Price, Matthew A and McEwen, Jason D},
	journal={arXiv preprint arXiv:2209.13603},
	year={2022}
}

@inproceedings{wang2021memory,
  title={Memory-efficient learning for high-dimensional {MRI} reconstruction},
  author={Wang, Ke and Kellman, Michael and Sandino, Christopher M and Zhang, Kevin and Vasanawala, Shreyas S and Tamir, Jonathan I and Yu, Stella X and Lustig, Michael},
  booktitle={International Conference on Medical Image Computing and Computer-Assisted Intervention},
  pages={461--470},
  year={2021},
  organization={Springer}
}

@inproceedings{wong2023fnoseg3d,
  title={{FNOSeg3D}: resolution-robust {3D} image segmentation with {Fourier} Neural Operator},
  author={Wong, Ken CL and Wang, Hongzhi and Syeda-Mahmood, Tanveer},
  booktitle={2023 IEEE 20th International Symposium on Biomedical Imaging (ISBI)},
  pages={1--5},
  year={2023},
  organization={IEEE}
}

@article{desai2022skm,
  title={{SKM-TEA}: A Dataset for Accelerated {MRI} Reconstruction with Dense Image Labels for Quantitative Clinical Evaluation},
  author={Desai, Arjun D and Schmidt, Andrew M and Rubin, Elka B and Sandino, Christopher M and Black, Marianne S and Mazzoli, Valentina and Stevens, Kathryn J and Boutin, Robert and R{\'e}, Christopher and Gold, Garry E and others},
  journal={arXiv preprint arXiv:2203.06823},
  year={2022}
}

@article{davoodi2025imaging,
  title={Imaging-guided deep tissue in vivo sound printing},
  author={Davoodi, Elham and Li, Jiahong and Ma, Xiaotian and Najafabadi, Alireza Hasani and Yoo, Jounghyun and Lu, Gengxi and Sani, Ehsan Shirzaei and Lee, Sunho and Montazerian, Hossein and Kim, Gwangmook and others},
  journal={Science},
  volume={388},
  number={6747},
  pages={616--623},
  year={2025},
  publisher={American Association for the Advancement of Science}
}

@article{rau20183d,
  title={{3D} functional ultrasound imaging of pigeons},
  author={Rau, Richard and Kruizinga, Pieter and Mastik, Frits and Belau, Markus and de Jong, Nico and Bosch, Johannes G and Scheffer, Wolfgang and Maret, Georg},
  journal={Neuroimage},
  volume={183},
  pages={469--477},
  year={2018},
  publisher={Elsevier}
}

@article{li2024pino,
  title={Physics-informed neural operator for learning partial differential equations},
  author={Li, Zongyi and Zheng, Hongkai and Kovachki, Nikola and Jin, David and Chen, Haoxuan and Liu, Burigede and Azizzadenesheli, Kamyar and Anandkumar, Anima},
  journal={ACM/JMS Journal of Data Science},
  volume={1},
  number={3},
  pages={1--27},
  year={2024},
  publisher={ACM New York, NY}
}

@article{lin2025mGNO,
title={Enabling Automatic Differentiation with Mollified Graph Neural Operators},
author={Ryan Y. Lin and Julius Berner and Valentin Duruisseaux and David Pitt and Daniel Leibovici and Jean Kossaifi and Kamyar Azizzadenesheli and Anima Anandkumar},
journal={Transactions on Machine Learning Research},
issn={2835-8856},
year={2025},
url={https://openreview.net/forum?id=CGoR1hFAGr},
}

@misc{ganeshram2025fcpinohighprecisionphysicsinformed,
      title={{FC-PINO}: High Precision Physics-Informed Neural Operators via {Fourier} Continuation}, 
      author={Adarsh Ganeshram and Haydn Maust and Valentin Duruisseaux and Zongyi Li and Yixuan Wang and Daniel Leibovici and Oscar Bruno and Thomas Hou and Anima Anandkumar},
      year={2025},
      eprint={2211.15960},
      archivePrefix={arXiv},
      primaryClass={cs.LG},
}

@article{berner2025principled,
  title={Principled Approaches for Extending Neural Architectures to Function Spaces for Operator Learning},
  author={Berner, Julius and Liu-Schiaffini, Miguel and Kossaifi, Jean and Duruisseaux, Valentin and Bonev, Boris and Azizzadenesheli, Kamyar and Anandkumar, Anima},
  journal={arXiv preprint arXiv:2506.10973},
  year={2025}
}

@misc{duruisseaux2025FNOGuide,
      title={Fourier Neural Operators Explained: A Practical Perspective}, 
      author={Valentin Duruisseaux and Jean Kossaifi and Anima Anandkumar},
      year={2025},
      eprint={2512.01421},
      archivePrefix={arXiv},
      primaryClass={cs.LG},
      url={https://arxiv.org/abs/2512.01421}, 
}

@article{Kovachki2021_2,
author = {Kovachki, Nikola and Lanthaler, Samuel and Mishra, Siddhartha},
title = {On Universal Approximation and Error Bounds for {F}ourier Neural Operators},
year = {2021},
publisher = {JMLR.org},
volume = {22},
number = {1},
issn = {1532-4435},
journal = {J. Mach. Learn. Res.},
articleno = {290},
}

@article{guo2024reconformer,
  title={{ReconFormer}: Accelerated {MRI} Reconstruction Using Recurrent Transformer},
  author={Guo, Pengfei and Mei, Yiqun and Zhou, Jinyuan and Jiang, Shanshan and Patel, Vishal M},
  journal={IEEE transactions on medical imaging},
  volume={43},
  number={1},
  pages={582--593},
  year={2023},
  publisher={IEEE}
}

@inproceedings{li2025lmo,
  title={{LMO}: Linear {Mamba} Operator for {MRI} Reconstruction},
  author={Li, Wei and Jiang, Jiawei and Wu, Jie and Yu, Kaihao and Zheng, Jianwei},
  booktitle={2025 IEEE/CVF Conference on Computer Vision and Pattern Recognition (CVPR)},
  pages={5112--5122},
  year={2025},
  organization={IEEE}
}
\clearpage
\newpage{}

\appendix

\section*{Appendix}

We provide details omitted in the main text in the appendix, including:
{
\setlength{\leftmargini}{2.4em}
\begin{itemize}\itemsep7pt


    \item \textbf{Appendix~\ref{app:dataset}: Dataset and Setup.} Preprocessing and coarse-to-fine resolution protocol for SKM-TEA, plus the multi-coil $k$-space simulation pipeline used for BraTS~2021.

    \item \textbf{Appendix~\ref{app:model}: KIR-NO Implementation.} Architecture, training hardware, hyperparameters, and compute for KIR-NO and the CNN baselines.

    \item \textbf{Appendix~\ref{app:gpu}: Memory Footprint.} Why full-resolution training is memory-bound, and how the cost of the DISCO operators scales with resolution.

    \item \textbf{Appendix~\ref{app:diffusion}: Diffusion Comparison.} Single-step and $10$-step comparisons against two diffusion baselines.

    \item \textbf{Appendix~\ref{app:ablations}: Additional Ablations and Experiments.} Additional baselines; robustness to measurement noise; tumor-ROI fidelity on BraTS; cross-scanner transfer (SKM-TEA\,$\to$\,fastMRI); the $k$-space head; basis function choice; kernel configuration and radius cutoff; number of cascades; and sampling-pattern robustness.

    \item \textbf{Appendix~\ref{app:broader_impacts}: Broader Impacts.} Potential benefits, risks, and the limits of clinical use.

\end{itemize}

}

\section{Dataset and Setup}
\label{app:dataset}

\textbf{SKM-TEA.} The SKM-TEA dataset~\citep{desai2022skm} is a multi-coil knee MRI dataset acquired with the qDESS sequence, with two echoes per sample; we use the first echo in all experiments. Scans use a $416 \times 512$ acquisition matrix (readout $\times$ phase encode; $0.38 \times 0.31$\,mm in-plane), $8$ or $16$ receive coils, and $80$ slices. The released $k$-space is zero-padded by the dataset to $512 \times 512$ in-plane and to twice the number of acquired slices, giving the $512 \times 512 \times 160$ volumes we start from.

We further interpolate the last dimension from $160$ to $320$, yielding volumes of $512 \times 512 \times 320$. We denote this as the full-resolution dataset. Downscaling by a factor of $2$ gives the half-resolution dataset, $256 \times 256 \times 160$, and a further factor of $2$ gives the quarter-resolution dataset, $128 \times 128 \times 80$. This enables a coarse-to-fine evaluation protocol: models trained at quarter resolution are evaluated zero-shot at half and full resolution, while models trained at half resolution are evaluated zero-shot at quarter and full resolution. At every resolution, the $k$-space data are formed by resizing the volume in \emph{image} space, followed by applying the FFT and the undersampling mask. We never interpolate in $k$-space, so the $k$-space data at each resolution are the exact transform of a clean image on that grid. At zero-shot evaluation, a learned model keeps the measurements of its training resolution; only the intermediate image is resampled to the evaluation grid (Section~\ref{sec:experiments}). No model sees evaluation-resolution examples during training.

\textbf{BraTS 2021.}
The BraTS~2021 dataset~\citep{brats2021} is a brain tumor segmentation benchmark containing pre-operative multimodal MRI volumes. We use the T1Gd modality with $100$ volumes for training per epoch and $20$ test volumes for evaluation. The original data are preprocessed NIfTI volumes at $240 \times 240 \times 155$ image-space resolution with isotropic $1\,\text{mm}^3$ voxel spacing.

Unlike SKM-TEA, BraTS provides only reconstructed magnitude images, so we synthesize multi-coil $k$-space. For each volume, we generate an $8$-coil sensitivity map on a virtual receive array, multiply the magnitude image by each coil profile to obtain complex coil images, and apply an orthonormal 3D FFT to produce simulated $k$-space. We then apply a $4\times$ Gaussian-density Cartesian undersampling mask, as for SKM-TEA.

To match the SKM-TEA coarse-to-fine protocol, we generate full-, half-, and quarter-resolution volumes of size $240 \times 240 \times 155$, $120 \times 120 \times 78$, and $60 \times 60 \times 39$, respectively, using trilinear downsampling in image space before the FFT. KIR-NO and the baselines are trained once per regime, at half ($120 \times 120 \times 78$) or at quarter ($60 \times 60 \times 39$) resolution, and evaluated at all three resolutions (Table~\ref{tab:brats}). The tumor-ROI study in Appendix~\ref{app:roi} uses the quarter-trained models. Because BraTS starts from coil-combined magnitude images rather than raw multi-coil acquisitions, the simulated setting lacks the acquisition diversity of SKM-TEA. As for SKM-TEA, the coil sensitivity maps are provided to all models as a known operator.

\section{KIR-NO for 3D MRI Reconstruction from Undersampled Measurements} \label{app:model}

KIR-NO is trained on a single NVIDIA B200 with a batch size of one volume. Each neural operator ``NO'' layer in Fig.~\ref{fig:arch} is a U-shaped DISCO Neural Operator~\citep{jatyani2025nomri}.

We evaluate two DISCO basis functions, the 3D Morlet and 3D piecewise-linear bases (Section~\ref{sec:methods_basis}). On SKM-TEA, KIR-NO uses the piecewise-linear basis with a kernel configuration $5\times5\times5$ with a radius cutoff of $0.08$ for the quarter-resolution model and $0.04$ for the half-resolution model. Unless stated otherwise, KIR-NO uses a single cascade with 18 channels along with 4 encoder, decoder and pooling layers for SKM-TEA and 2 encoder, decoder and pooling layers for BraTS. Ablations over the basis function and the number of cascades are in Appendix~\ref{app:basis} (Table~\ref{tab:basis_ablation}) and Appendix~\ref{app:deepcas} (Table~\ref{tab:cascades_cnn_kno}).

The 3D CNN baseline is E2E-VarNet~\citep{e2evarnet} with 2D convolutions replaced by 3D convolutions; it uses the provided coil sensitivity maps in place of the learned sensitivity module (Section~\ref{sec:experiments}).

All models are trained with the Adam optimizer, and hyperparameters were selected on validation performance. On SKM-TEA, all models are trained for $200$ epochs at quarter resolution and $120$ epochs at half resolution; on BraTS, all learned models are trained for $120$ epochs. On BraTS, both KIR-NO models use a radius cutoff of $0.302$ and a kernel configuration of $7\times7\times7$.

\textbf{Compute requirements.}
Training the main SKM-TEA half-resolution KIR-NO model ($120$ epochs) takes
approximately $43$ GPU-hours on a single NVIDIA B200; the quarter-resolution model ($200$ epochs) takes approximately $5.4$
GPU-hours. 

%

\section{Memory Footprint and Coarse-to-Fine Deployability} \label{app:gpu}

Full-volume 3D reconstruction is memory-bound at training time: a single full-resolution ($512\times512\times320$) training step exceeded the memory of a GH200 GPU for \emph{both} the 3D~E2E-VarNet CNN and KIR-NO, and KIR-NO's full-resolution inference alone needs $72.2$\,GiB before any activations are stored for back-propagation. Native full-resolution training would require model parallelism, gradient checkpointing, or resolution downscaling. The coarse-to-fine protocol avoids this: models are trained where back-propagation is memory-feasible (quarter or half resolution) and deployed zero-shot at full resolution (Tables~\ref{tab:zero_shot_sr}--\ref{tab:brats}).

The learned DISCO coefficients do not depend on resolution, but each kernel is defined over a fixed physical radius $r_c$. On a finer grid, this radius contains more voxels, so the discrete stencil obtained by sampling the kernel has more entries. The inference cost per output voxel of the image-space operators therefore grows with the evaluation resolution.

Full-resolution \emph{inference} is feasible on high-memory GPUs. Because the image-space DISCO operators are local, peak inference memory could in principle be reduced further by exact tiled execution with coil-streamed data consistency, trading runtime for memory.

\section{Diffusion Comparison} \label{app:diffusion}

We compare KIR-NO with two diffusion-based reconstruction methods~\citep{krainovic2024resolution,jalal2021csgm} under matched inference budgets. Since diffusion models typically require many sampling steps at test time, we evaluate two regimes: a \emph{single-forward} setting, in which KIR-NO is evaluated with one forward pass and each diffusion baseline with one reverse-sampling step, and a \emph{multi-step iterative} setting, in which both methods are allowed $10$ inference steps. For KIR-NO, the $10$ steps are $10$ unrolled image-space cascades: a separately trained $10$-cascade model, evaluated in one forward pass. For the diffusion baselines, the steps are $10$ score-based reverse-diffusion steps.

Table~\ref{tab:iterative} reports the quantitative comparison. In the single-forward setting, KIR-NO achieves $33.13$\,dB at quarter resolution and $33.69$\,dB at half resolution, ahead of the strongest one-step diffusion baseline by $+6.14$\,dB and $+1.27$\,dB respectively, and ahead on SSIM and NMSE at both resolutions. The $10$-step setting is less uniform. At quarter resolution KIR-NO reaches $35.94$\,dB and stays $+6.21$\,dB ahead of the best $10$-step diffusion baseline, and that is $2.81$\,dB above its own single-forward output and better on SSIM and NMSE as well, so the additional cascades help at this resolution. At half resolution KIR-NO improves to $34.25$\,dB, a gain of $+0.56$\,dB over its single forward, while the 3D physics-based baseline reaches $37.75$\,dB and leads on all three metrics, overtaking KIR-NO by $3.50$\,dB. 

\begin{table}[h] 
  \centering
 \caption{Iterative reconstruction on SKM-TEA at quarter and half resolution
($4\times$ acceleration; KIR-NO trained at quarter resolution, diffusion priors at half resolution).
\emph{Top:} single-forward setting, where each method performs one inference
step. \emph{Bottom:} multi-step setting with $10$ inference steps: KIR-NO is a
$10$-cascade model evaluated in one forward pass, while the diffusion baselines take
$10$ score-based reverse-diffusion steps. Results at the diffusion baselines' full
sampling trajectories are given in the text below.}

  \label{tab:iterative}
  \footnotesize
  \resizebox{\linewidth}{!}{
  \begin{tabular}{l ccc c ccc}
  \toprule
  & \multicolumn{3}{c}{\textbf{Quarter-Res (128)}} & & \multicolumn{3}{c}{\textbf{Half-Res (256)}} \\
  \cmidrule(lr){2-4} \cmidrule(lr){6-8}
  Method & PSNR$\uparrow$ & SSIM$\uparrow$ & NMSE$\downarrow$ & &
           PSNR$\uparrow$ & SSIM$\uparrow$ & NMSE$\downarrow$ \\
  \midrule
  \multicolumn{8}{l}{\textit{Single forward}} \\
  \midrule
  Diffusion-3D Physics \citep{krainovic2024resolution} & $26.99_{\pm 1.33}$ & $0.716_{\pm 0.035}$ & $0.0982_{\pm 0.0214}$ & & $32.42_{\pm 1.57}$ & $0.827_{\pm 0.027}$ & $0.0341_{\pm 0.0104}$ \\
  Diffusion-CSGM \citep{jalal2021csgm} & $26.61_{\pm 1.31}$ & $0.701_{\pm 0.036}$ & $0.1069_{\pm 0.0222}$ & & $29.89_{\pm 1.50}$ & $0.799_{\pm 0.031}$ & $0.0603_{\pm 0.0146}$ \\
    KIR-NO (Ours)                                       & $\mathbf{33.13}_{\pm 1.44}$ & $\mathbf{0.894}_{\pm 0.020}$ & $\mathbf{0.0242}_{\pm 0.0069}$ & & $\mathbf{33.69}_{\pm 1.45}$ & $\mathbf{0.872}_{\pm 0.021}$ & $\mathbf{0.0251}_{\pm 0.0060}$ \\
  \midrule
  \multicolumn{8}{l}{\textit{Multi-step iterative (10)}} \\
  \midrule
  Diffusion-3D Physics \citep{krainovic2024resolution} & $29.73_{\pm 1.41}$ & $0.787_{\pm 0.032}$ & $0.0530_{\pm 0.0150}$ & & $\mathbf{37.75}_{\pm 1.51}$ & $\mathbf{0.921}_{\pm 0.014}$ & $\mathbf{0.0100}_{\pm 0.0029}$ \\
  Diffusion-CSGM \citep{jalal2021csgm} & $28.25_{\pm 1.43}$ & $0.751_{\pm 0.034}$ & $0.0742_{\pm 0.0192}$ & & $34.82_{\pm 1.76}$ & $0.885_{\pm 0.024}$ & $0.0200_{\pm 0.0072}$ \\
  \textbf{KIR-NO (Ours)} & $\mathbf{35.94}_{\pm 1.44}$ & $\mathbf{0.933}_{\pm 0.011}$ & $\mathbf{0.0127}_{\pm 0.0037}$ & & $34.25_{\pm 1.45}$ & $0.899_{\pm 0.018}$ & $0.0221_{\pm 0.0056}$ \\
  \bottomrule
  \end{tabular}}
\end{table}

\paragraph{Diffusion baselines at their original sampling budgets.}
The comparison above equalizes the number of inference steps. We also evaluate both diffusion baselines at the sampling
budgets their authors specify: $200$ reverse steps for the 3D physics-based model and $1200$
annealed Langevin steps for CSGM. Table~\ref{tab:diffusion_cost} reports per-volume
reconstruction time on BraTS $8$-coil volumes at $4\times$ acceleration, on a single idle B200.

\begin{table}[h!]
    \centering
    \caption{Per-volume inference time on BraTS 8-coil volumes at $4\times$, for the diffusion baselines and KIR-NO.}
    \label{tab:diffusion_cost}
    \footnotesize
    \begin{tabular}{lcc}
        \toprule
        \textbf{Method} & \textbf{QR $60\times60\times39$} & \textbf{HR $120\times120\times78$} \\
        \midrule
        KIR-NO (single cascade) & $0.157$\,s & $3.22$\,s \\
        Diffusion-3D Physics (single step) & $0.079$\,s & $0.096$\,s \\
        Diffusion-CSGM (2D, single step, batched) & $0.060$\,s & $0.082$\,s \\
        Diffusion-3D Physics ($200$ steps) & $10.1$\,s & $13.1$\,s \\
        Diffusion-CSGM (2D, $1200$ steps) & $35.6$\,s & $58.9$\,s \\
        \textbf{KIR-NO (10 cascades)} & $1.14$\,s & $17.8$\,s \\
        \bottomrule
    \end{tabular}
\end{table}

A single KIR-NO cascade reconstructs a quarter-resolution volume in $0.157$\,s
(Table~\ref{tab:diffusion_cost}); the 3D physics-based baseline
takes $64\times$ more time per volume at that resolution with full sampling. The margin is resolution-dependent and much
smaller at half resolution, because the operator's kernel support scales with the grid while a
sampling step's does not. All times are device-resident reconstruction.

\paragraph{Accuracy at the full sampling trajectory.}
Given enough steps, both diffusion baselines are accurate. Sampling CSGM along its dense
trajectory ($1200$ steps) on SKM-TEA gives $36.88$\,dB / $0.9357$ / $0.0101$ at
quarter resolution and $44.47$\,dB / $0.9829$ / $0.0021$ at half resolution; the 3D
physics-based model reaches $35.04$\,dB / $0.922$ and $38.46$\,dB / $0.933$ on the same grids. The 3D
physics-based model already exceeds KIR-NO at half resolution with $10$ steps
(Table~\ref{tab:iterative}); along its full trajectory it does so at quarter resolution as well,
and CSGM exceeds KIR-NO at both resolutions.
KIR-NO's single forward pass is $1.9$--$3.8$\,dB below these values at quarter resolution and
$4.8$--$10.8$\,dB below at half resolution; its advantage is inference time
(Table~\ref{tab:diffusion_cost}).

\section{Additional Ablations and Experiments}\label{app:ablations}

This section reports additional baselines, robustness tests (measurement noise, cross-scanner transfer, sampling pattern), tumor-ROI metrics, and ablations of the $k$-space head, basis function, kernel configuration and radius cutoff, and number of cascades.

\subsection{Additional Baseline Comparison}\label{app:recent_baselines}

Beyond the classical and CNN baselines in the main text, we compare against three additional 
recent learned reconstruction methods: LMO (Linear Mamba Operator)~\citep{li2025lmo}, the
recurrent transformer ReconFormer~\citep{guo2024reconformer}, and the 2D neural-operator
CS-MRI model of~\citet{jatyani2025nomri} (``2D-NO''). All learned methods are trained on
quarter-resolution SKM-TEA ($128\!\times\!128\!\times\!80$, $4\times$) and evaluated across
resolutions with fixed weights, so half and full resolution are out-of-domain for every
method. For these 2D (slice-wise) baselines we report PSNR only. KIR-NO outperforms LMO and
2D-NO at every resolution, whereas ReconFormer attains higher PSNR than KIR-NO at all three
resolutions.

\begin{table}[h!]
    \centering
    \caption{Comparison with additional baselines on SKM-TEA
    ($4\times$), PSNR in dB. All learned models are trained at quarter resolution
    ($128\!\times\!128\!\times\!80$) and evaluated at quarter (in-domain), half, and full
    resolution with fixed weights.}
    \label{tab:recent_baselines}
    \footnotesize
    \begin{tabular}{lccc}
        \toprule
        \textbf{Method} & \textbf{Quarter (128)} & \textbf{Half (256)} & \textbf{Full (512)} \\
        \midrule
        LMO~\citep{li2025lmo} &{29.12} & {31.49} & {33.02} \\
        ReconFormer~\citep{guo2024reconformer}        & {34.16} & {36.80} & {39.05} \\
        2D-NO~\citep{jatyani2025nomri}                & {28.76} & {29.30} & {31.11} \\
        \textbf{KIR-NO (Ours)}                        & {33.13} & {33.69} & {35.52} \\
        \bottomrule
    \end{tabular}
\end{table}

\hfill 

\subsection{Robustness to Measurement Noise}\label{app:noise}

We stress-test robustness to acquisition noise by injecting complex Gaussian
noise (AWGN) at the \emph{sampled} $k$-space positions at a controlled measurement SNR and
evaluating at half resolution ($4\times$). The noise is applied at evaluation time only: both
models are the clean-trained half-resolution models of Tables~\ref{tab:zero_shot_sr}
and~\ref{tab:accel_rates}, so the comparison is \emph{zero-shot}
for the noisy data. KIR-NO's advantage narrows monotonically as the measurement SNR
falls: $+2.23$\,dB with no noise, $+1.81$\,dB at $20$\,dB SNR and
$+0.79$\,dB at $14$\,dB. On SSIM the CNN already leads at
$14$\,dB SNR noise ($0.8624$ versus $0.8409$), where KIR-NO still leads on PSNR. KIR-NO thus leads in PSNR
at every tested SNR but degrades faster than the CNN as noise increases ($-2.14$\,dB from clean
to $14$\,dB SNR, against the CNN's $-0.70$\,dB); see Table~\ref{tab:noise}.

\begin{table}[h!]
    \centering
    \caption{Robustness to measurement noise on SKM-TEA half resolution
    ($4\times$), PSNR in dB; both models are trained on
    clean half-resolution evaluated zero-shot}
    \label{tab:noise}
    \footnotesize
    \begin{tabular}{lccc}
        \toprule
        \textbf{Measurement SNR (dB)} & \textbf{No noise} & \textbf{20} & \textbf{14} \\
        \midrule
        3D E2E-VarNet         & $34.31_{\pm 1.73}$ & $34.14_{\pm 1.70}$ & $33.61_{\pm 1.62}$ \\
        \textbf{KIR-NO (Ours)} & $\mathbf{36.54}_{\pm 1.60}$ & $\mathbf{35.95}_{\pm 1.51}$ & $\mathbf{34.40}_{\pm 1.38}$ \\
        \bottomrule
    \end{tabular}
\end{table}

\clearpage

\subsection{Clinical ROI Metrics (Tumor Fidelity)}\label{app:roi}

Beyond whole-image PSNR/SSIM/NMSE, we report tumor-region metrics (Table~\ref{tab:roi}) at quarter resolution ($60\times60\times39$, $4\times$, $8$ coils) for KIR-NO and the 3D E2E-VarNet, both trained at quarter resolution, on the $20$ test volumes of Table~\ref{tab:brats}. The ROI is defined volumetrically for the whole tumor: comprising the necrotic core, edema and enhancing tumor. Metrics are reported as PSNR / SSIM / NMSE.

\begin{table}[h!]
    \centering
    \caption{BraTS tumor-region fidelity at quarter resolution ($4\times$, $8$ coils). Both learned models are trained at quarter resolution.}
    \label{tab:roi}
    \resizebox{\linewidth}{!}{
    \begin{tabular}{l ccc ccc}
        \toprule
        & \multicolumn{3}{c}{\textbf{Whole image}} & \multicolumn{3}{c}{\textbf{Tumor ROI}} \\
        \cmidrule(lr){2-4}\cmidrule(lr){5-7}
        \textbf{Method} & PSNR$\uparrow$ & SSIM$\uparrow$ & NMSE$\downarrow$ & PSNR$\uparrow$ & SSIM$\uparrow$ & NMSE$\downarrow$ \\
        \midrule
        Zero-Filled & $30.59_{\pm 1.45}$ & $0.756_{\pm 0.028}$ & $0.0663_{\pm 0.0135}$ & $23.95_{\pm 2.65}$ & $0.750_{\pm 0.047}$ & $0.0427_{\pm 0.0216}$ \\
        3D E2E-VarNet (CNN) & $37.73_{\pm 1.47}$ & $\mathbf{0.964}_{\pm 0.004}$ & $0.0129_{\pm 0.0031}$ & $28.39_{\pm 1.64}$ & $0.908_{\pm 0.024}$ & $0.0142_{\pm 0.0042}$ \\
        \textbf{KIR-NO (Ours)} & $\mathbf{40.14}_{\pm 1.48}$ & $0.958_{\pm 0.008}$ & $\mathbf{0.0075}_{\pm 0.0022}$ & $\mathbf{31.02}_{\pm 1.98}$ & $\mathbf{0.947}_{\pm 0.014}$ & $\mathbf{0.0081}_{\pm 0.0035}$ \\
        \bottomrule
    \end{tabular}}
\end{table}

We observe that KIR-NO outperforms baseline methods for tumor ROI regions.


\hfill 

\subsection{Cross-Scanner Transfer (SKM-TEA $\to$ fastMRI Knee)}\label{app:crossscanner}

We test a genuinely different acquisition: models trained on quarter-resolution SKM-TEA and applied to fastMRI knee~\citep{zbontar2018fastMRI}, which differs in scanner, coil array, protocol, and SNR. In the few-shot setting both models are fine-tuned for $15$ epochs on $100$ fastMRI knee volumes. Evaluation is on $30$ unseen fastMRI whole-knee volumes at native resolution (Table~\ref{tab:crossscanner}).

\begin{table}[h!]
    \centering
    \caption{Cross-scanner transfer to fastMRI knee (PSNR / SSIM), on $30$ unseen whole-knee volumes at native resolution ($0.44\times0.44\times3$\,mm, $4\times$). Both models are trained on quarter-resolution SKM-TEA; few-shot fine-tunes on $100$ fastMRI volumes for $15$ epochs.}
    \label{tab:crossscanner}
    \footnotesize
    \begin{tabular}{lcc}
        \toprule
        \textbf{Evaluation on fastMRI knee} & \textbf{Zero-shot} & \textbf{Few-shot} \\
        \midrule
        3D E2E-VarNet (CNN) & 26.33 / 0.711 & 31.10 / 0.824 \\
        \textbf{KIR-NO (Ours)} & \textbf{29.12 / 0.769} & \textbf{36.17 / 0.929} \\
        \bottomrule
    \end{tabular}
\end{table}

KIR-NO leads in both settings, and its margin over the VarNet widens with fine-tuning ($+2.79\to+5.07$\,dB).

\hfill 

\subsection{Effect of the $k$-space Head ($k$NO)}\label{app:kno}

We ablate the $k$-space neural-operator head by removing it, leaving only the image-space DISCO cascades (\emph{KIR-NO w/o $k$NO}). Both variants are trained at quarter resolution and evaluated at quarter, half, and full resolution (Table~\ref{tab:kno}). The $k$NO head is key to KIR-NO's high-resolution performance: without it, PSNR \emph{drops} as evaluation resolution grows ($31.29\to30.22\to28.89$\,dB), whereas the full model \emph{improves} ($33.13\to33.69\to35.52$\,dB), widening the gap to $+6.6$\,dB at full resolution. The image-space cascades alone therefore do not account for KIR-NO's high-resolution gains.

\begin{table}[h!]
    \centering
    \caption{{Effect of the $k$-space head. Both variants are trained at quarter resolution SKM-TEA.}}
    \label{tab:kno}
    \footnotesize
    \resizebox{\linewidth}{!}{
    \begin{tabular}{l ccc ccc ccc}
    \toprule
    & \multicolumn{3}{c}{\textbf{Quarter (128)}}
    & \multicolumn{3}{c}{\textbf{Half (256)}}
    & \multicolumn{3}{c}{\textbf{Full (512)}} \\
    \cmidrule(lr){2-4}\cmidrule(lr){5-7}\cmidrule(lr){8-10}
    \textbf{Method} & PSNR & SSIM & NMSE & PSNR & SSIM & NMSE & PSNR & SSIM & NMSE \\
    \midrule
    KIR-NO w/o $k$NO & $31.29_{\pm 1.47}$ & $0.865_{\pm 0.024}$ & $0.0370_{\pm 0.0104}$ & $30.22_{\pm 1.36}$ & $0.823_{\pm 0.028}$ & $0.0554_{\pm 0.0107}$ & $28.89_{\pm 1.18}$ & $0.832_{\pm 0.020}$ & $0.0850_{\pm 0.0117}$ \\
    \textbf{KIR-NO (ours)} & $\mathbf{33.13}_{\pm 1.44}$ & $\mathbf{0.894}_{\pm 0.020}$ & $\mathbf{0.0242}_{\pm 0.0069}$ & $\mathbf{33.69}_{\pm 1.45}$ & $\mathbf{0.872}_{\pm 0.021}$ & $\mathbf{0.0251}_{\pm 0.0060}$ & $\mathbf{35.52}_{\pm 1.22}$ & $\mathbf{0.908}_{\pm 0.014}$ & $\mathbf{0.0187}_{\pm 0.0038}$ \\
    \bottomrule
    \end{tabular}}
\end{table}

\hfill 

\subsection{Basis Function Choice}\label{app:basis}

We compare two DISCO bases: piecewise-linear3d and Morlet3d. Table~\ref{tab:basis_ablation} isolates the basis: both KIR-NO models are trained and evaluated at SKM-TEA quarter resolution ($128{\times}128{\times}80$) with the \emph{identical} kernel configuration ($5, 5, 5$), radius cutoff ($0.08$) and 200-epoch training protocol of Table~\ref{tab:zero_shot_sr}, and differ only in the basis. Piecewise-linear3d outperforms Morlet3d by $+2.11\,\mathrm{dB}$ ($33.13$ vs $31.02\,\mathrm{dB}$). Crucially, the Morlet kernel carries \emph{more} learnable coefficients than PL, so the piecewise-linear advantage reflects the basis's inductive bias.

\begin{table}[h!]  
    \centering
    \captionsetup{singlelinecheck=false}   
    \caption{{Basis-function ablation on SKM-TEA quarter-resolution reconstruction.}}
    \label{tab:basis_ablation}
    \renewcommand{\arraystretch}{1.15}
    \setlength{\tabcolsep}{8pt}
    \begin{tabular}{llccc}
        \toprule
        \textbf{Model} & \textbf{Basis} & \textbf{PSNR $\uparrow$} & \textbf{SSIM $\uparrow$} & \textbf{NMSE $\downarrow$} \\
        \midrule
        KIR-NO & Morlet & $31.02_{\pm 1.41}$ & $0.8320_{\pm 0.027}$ & $0.0393_{\pm 0.0106}$ \\
    {KIR-NO} & Piecewise-linear & $\mathbf{33.13}_{\pm 1.44}$ & $\mathbf{0.8943}_{\pm 0.020}$ & $\mathbf{0.0242}_{\pm 0.0069}$ \\
        \bottomrule
    \end{tabular}  
\end{table}


\hfill 

\subsection{Kernel Configuration and Radius Cutoff}\label{app:hyperparam}

Table~\ref{tab:hyperparam} ablates the kernel configuration and radius cutoff for KIR-NO using the piecewise-linear 3D basis. Two trends are apparent. First, a radius cutoff of $0.08$  outperforms other settings like $0.05, 0.13$  across all kernel configurations, suggesting that a medium-large DISCO receptive field is important for reconstruction quality. Second, the $7 \times 7 \times 7$ kernel with cutoff $0.05$ achieves the best performance, reaching $33$\,dB and slightly outperforming other kernel configurations. Additionally, reconstruction performance is relatively robust to slight changes in hyperparameters, like kernel configurations and convolution size $\mathbf{r_c}$. 

\begin{table}[h!] 
    \centering
    \captionsetup{singlelinecheck=false}   
    \caption{Kernel configuration  and radius-cutoff sweep for KIR-NO on SKM-TEA Dataset}
    \label{tab:hyperparam}
    \footnotesize
    \renewcommand{\arraystretch}{1.15}
    \resizebox{\linewidth}{!}{
    \begin{tabular}{lc ccc ccc}
        \toprule
        & & \multicolumn{3}{c}{\textbf{Quarter Res (128)}}
        & \multicolumn{3}{c}{\textbf{Half Res (256)}} \\
        \cmidrule(lr){3-5} \cmidrule(lr){6-8}
        \textbf{Kernel Config} & $\mathbf{r_c}$ & PSNR$\uparrow$ & SSIM$\uparrow$ & NMSE$\downarrow$
                                             & PSNR$\uparrow$ & SSIM$\uparrow$ & NMSE$\downarrow$ \\
        \midrule
        $3\times3\times3$ & $0.05$ & $32.74_{\pm 1.42}$ & $0.886_{\pm 0.021}$ & $0.0265_{\pm 0.0073}$ & $33.38_{\pm 1.47}$ & $0.859_{\pm 0.023}$ & $0.0270_{\pm 0.0066}$ \\
        $5\times5\times5$ & $0.05$ & $32.86_{\pm 1.43}$ & $0.888_{\pm 0.021}$ & $0.0258_{\pm 0.0073}$ & $33.02_{\pm 1.41}$ & $0.851_{\pm 0.023}$ & $0.0292_{\pm 0.0064}$ \\
        $7\times7\times7$ & $0.05$ & $32.94_{\pm 1.42}$ & $0.890_{\pm 0.020}$ & $0.0253_{\pm 0.0069}$ & $33.03_{\pm 1.38}$ & $0.856_{\pm 0.022}$ & $0.0289_{\pm 0.0054}$ \\
        \midrule
        $3\times3\times3$ & $0.08$ & $31.76_{\pm 1.47}$ & $0.860_{\pm 0.026}$ & $0.0332_{\pm 0.0094}$ & $32.77_{\pm 1.40}$ & $0.853_{\pm 0.023}$ & $0.0308_{\pm 0.0064}$ \\
        $5\times5\times5$ & $0.08$ & $33.13_{\pm 1.44}$ & $\mathbf{0.894}_{\pm 0.020}$ & $0.0242_{\pm 0.0069}$ & $33.69_{\pm 1.45}$ & $0.872_{\pm 0.021}$ & $0.0251_{\pm 0.0060}$ \\
        $7\times7\times7$ & $0.08$ & $\mathbf{33.15}_{\pm 1.45}$ & $\mathbf{0.894}_{\pm 0.020}$ & $\mathbf{0.0241}_{\pm 0.0070}$ & $\mathbf{33.89}_{\pm 1.42}$ & $\mathbf{0.877}_{\pm 0.020}$ & $\mathbf{0.0239}_{\pm 0.0055}$ \\
        \midrule
        $3\times3\times3$ & $0.13$ & $30.69_{\pm 1.37}$ & $0.811_{\pm 0.027}$ & $0.0422_{\pm 0.0106}$ & $32.90_{\pm 1.38}$ & $0.826_{\pm 0.025}$ & $0.0299_{\pm 0.0063}$ \\
        $5\times5\times5$ & $0.13$ & $32.69_{\pm 1.45}$ & $0.882_{\pm 0.022}$ & $0.0268_{\pm 0.0078}$ & $33.76_{\pm 1.41}$ & $0.873_{\pm 0.021}$ & $0.0247_{\pm 0.0058}$ \\
        $7\times7\times7$ & $0.13$ & $33.00_{\pm 1.45}$ & $0.889_{\pm 0.022}$ & $0.0250_{\pm 0.0073}$ & $33.16_{\pm 1.44}$ & $0.846_{\pm 0.026}$ & $0.0282_{\pm 0.0064}$ \\
        \bottomrule
    \end{tabular}}
\end{table}


\hfill 

\subsection{Deeper Cascades with Gradient Checkpointing}\label{app:deepcas}

KIR-NO uses a single image-space cascade by default for memory efficiency and for a controlled comparison, but gradient checkpointing enables deeper unrolling. Table~\ref{tab:cascades_cnn_kno} reports quarter-resolution ($4\times$) reconstruction as the number of cascades increases for both the 3D CNN and KIR-NO with checkpointing. KIR-NO remains ahead at every matched cascade count (e.g., $35.25\,\mathrm{dB}$ / $0.931$ SSIM versus $31.96\,\mathrm{dB}$ / $0.873$ SSIM at three cascades), confirming that the advantage is not an artifact of the single-cascade default. Both models gain most from the first increase in depth; KIR-NO saturates thereafter, while the CNN regresses slightly at five cascades.



\begin{table}[htb]
    \centering
    \captionsetup{singlelinecheck=false}   
    \caption{Cascades ablation on SKM-TEA Quarter-res ($4\times$).}
    \label{tab:cascades_cnn_kno}
    \footnotesize
    \resizebox{0.99\linewidth}{!}{
    \begin{tabular}{l ccc ccc}
    \toprule
    & \multicolumn{3}{c}{\textbf{E2E-VarNet3D (CNN)}}
    & \multicolumn{3}{c}{\textbf{KIR-NO (Ours)}} \\
    \cmidrule(lr){2-4}\cmidrule(lr){5-7}
    \textbf{Cascades} & PSNR & SSIM & NMSE & PSNR & SSIM & NMSE \\
    \midrule
    cas $= 1$ & $30.65_{\pm 1.50}$ & $0.8519_{\pm 0.030}$ & $0.0430_{\pm 0.0124}$ & $\mathbf{33.13}_{\pm 1.44}$ & $\mathbf{0.8943}_{\pm 0.020}$ & $\mathbf{0.0242}_{\pm 0.0069}$ \\
    cas $= 3$ & $31.96_{\pm 1.58}$ & $0.8732_{\pm 0.032}$ & $0.0319_{\pm 0.0096}$ & $\mathbf{35.25}_{\pm 1.46}$ & $\mathbf{0.9311}_{\pm 0.014}$ & $\mathbf{0.0149}_{\pm 0.0045}$ \\
    cas $= 5$ & $31.46_{\pm 1.39}$ & $0.8582_{\pm 0.026}$ & $0.0354_{\pm 0.0093}$ & $\mathbf{35.33}_{\pm 1.47}$ & $\mathbf{0.9314}_{\pm 0.013}$ & $\mathbf{0.0147}_{\pm 0.0044}$ \\
    \bottomrule
    \end{tabular}}
\end{table}

\hfill 

\subsection{Sampling-Pattern Robustness}\label{app:patterns}

We assess robustness to the $k$-space sampling pattern, complementing the Gaussian-density Cartesian mask used in the main experiments. Table~\ref{tab:pattern_ablation} reports quarter-resolution reconstruction under four patterns --- Gaussian (paper default), equispaced, radial, and 2D Poisson-disc. 

All learned models use a single cascade ($\text{cas}{=}1$) and are retrained per pattern at quarter resolution ($128{\times}128{\times}80$). For each of the other three patterns, a $4\times$ acceleration mask is applied to all training and evaluation volumes. KIR-NO attains the best PSNR, SSIM and NMSE under every pattern, with a PSNR margin over the strongest CNN ranging from $+1.5$\,dB (radial) to $+2.9$\,dB (Poisson); the Poisson-disc pattern yields the highest KIR-NO reconstruction overall ($35.30$\,dB / $0.922$ SSIM). These results indicate that the gains reported in the main text are not specific to the Gaussian mask.

\begin{table}[htb]
    \centering
    \captionsetup{singlelinecheck=false}   
    \caption{Sampling-pattern ablation on SKM-TEA Quarter-res.}
    \label{tab:pattern_ablation}
    \footnotesize
    \resizebox{\linewidth}{!}{
    \begin{tabular}{l ccc ccc ccc ccc ccc}
    \toprule
    & \multicolumn{3}{c}{\textbf{Zero-Filled}}
    & \multicolumn{3}{c}{\textbf{$\ell_1$-Wavelet}}
    & \multicolumn{3}{c}{\textbf{2D E2E-VarNet}}
    & \multicolumn{3}{c}{\textbf{3D E2E-VarNet}}
    & \multicolumn{3}{c}{\textbf{KIR-NO (Ours)}} \\
    \cmidrule(lr){2-4}\cmidrule(lr){5-7}\cmidrule(lr){8-10}\cmidrule(lr){11-13}\cmidrule(lr){14-16}
    \textbf{Pattern} & PSNR & SSIM & NMSE & PSNR & SSIM & NMSE & PSNR & SSIM & NMSE & PSNR & SSIM & NMSE & PSNR & SSIM & NMSE \\
    \midrule
    Gaussian (paper default) & $26.59_{\pm 1.31}$ & $0.700_{\pm 0.036}$ & $0.1074_{\pm 0.0222}$ & $26.62_{\pm 1.31}$ & $0.701_{\pm 0.036}$ & $0.1067_{\pm 0.0221}$ & $31.45_{\pm 1.46}$ & $0.874_{\pm 0.027}$ & $0.0357_{\pm 0.0099}$ & $30.65_{\pm 1.50}$ & $0.852_{\pm 0.030}$ & $0.0430_{\pm 0.0124}$ & $\mathbf{33.13}_{\pm 1.44}$ & $\mathbf{0.894}_{\pm 0.020}$ & $\mathbf{0.0242}_{\pm 0.0069}$ \\
    Equispaced & $26.44_{\pm 1.28}$ & $0.699_{\pm 0.032}$ & $0.1110_{\pm 0.0223}$ & $26.46_{\pm 1.29}$ & $0.699_{\pm 0.032}$ & $0.1105_{\pm 0.0221}$ & $31.91_{\pm 1.51}$ & $0.880_{\pm 0.026}$ & $0.0323_{\pm 0.0100}$ & $30.93_{\pm 1.58}$ & $0.855_{\pm 0.030}$ & $0.0407_{\pm 0.0132}$ & $\mathbf{33.48}_{\pm 1.53}$ & $\mathbf{0.895}_{\pm 0.019}$ & $\mathbf{0.0226}_{\pm 0.0074}$ \\
    Radial     & $27.71_{\pm 1.31}$ & $0.747_{\pm 0.032}$ & $0.0833_{\pm 0.0188}$ & $27.77_{\pm 1.32}$ & $0.748_{\pm 0.032}$ & $0.0823_{\pm 0.0186}$ & $32.34_{\pm 1.44}$ & $0.889_{\pm 0.023}$ & $0.0290_{\pm 0.0081}$ & $31.38_{\pm 1.53}$ & $0.862_{\pm 0.026}$ & $0.0364_{\pm 0.0108}$ & $\mathbf{33.87}_{\pm 1.43}$ & $\mathbf{0.912}_{\pm 0.017}$ & $\mathbf{0.0205}_{\pm 0.0059}$ \\
    Poisson & $25.48_{\pm 1.11}$ & $0.684_{\pm 0.027}$ & $0.1370_{\pm 0.0169}$ & $25.71_{\pm 1.11}$ & $0.687_{\pm 0.027}$ & $0.1301_{\pm 0.0166}$ & $32.40_{\pm 1.48}$ & $0.889_{\pm 0.023}$ & $0.0287_{\pm 0.0085}$ & $31.87_{\pm 1.55}$ & $0.876_{\pm 0.023}$ & $0.0326_{\pm 0.0099}$ & $\mathbf{35.30}_{\pm 1.44}$ & $\mathbf{0.922}_{\pm 0.014}$ & $\mathbf{0.0147}_{\pm 0.0043}$ \\
    \bottomrule
    \end{tabular}}
\end{table}

\normalcolor

\hfill \\

\section{Broader Impacts}
\label{app:broader_impacts}

\textbf{Positive impacts.}
Accelerated MRI reconstruction reduces scan times, lowers patient discomfort
(particularly for pediatric, claustrophobic, and trauma patients), and
increases scanner throughput, which can broaden access to high-quality
imaging in resource-constrained healthcare settings. Because KIR-NO can be
trained on memory-feasible low-resolution volumes and evaluated at higher
resolution, it lowers the GPU memory needed to train 3D MRI reconstruction
models.

\textbf{Potential negative impacts and mitigations.}
Learned MRI reconstruction methods can exhibit failure modes that are
visually subtle, including hallucinated structures or suppressed pathology
on out-of-distribution scans (e.g., unseen scanner vendors, contrasts,
anatomies, or pathology). We emphasize that KIR-NO is presented as a
research contribution and is \emph{not} validated for clinical deployment.
Any clinical use would require prospective validation, regulatory review,
integration of uncertainty quantification, and a fallback to conventional
reconstruction in out-of-distribution regimes. Our zero-shot evaluation
explicitly tests cross-resolution generalization but does not address
distribution shifts beyond the two anatomies and the sampling patterns
studied here.


\end{document}